\documentclass[12pt]{article}

\usepackage[body={17.5cm, 21cm},right=2cm]{geometry}

\usepackage{color}
\usepackage{graphicx}
\usepackage{epsf}
\usepackage{graphicx,epsfig}
\usepackage{amsmath}
\usepackage{amssymb}
\usepackage{epsfig}
\usepackage{cite}
\usepackage{color,colordvi}
\newcommand{\be}{\begin{eqnarray}}
\newcommand{\ee}{\end{eqnarray}}
\newcommand{\bi}{\begin{itemize}}
\newcommand{\ei}{\end{itemize}}

\newcommand{\bx}{{\bf{x}}}

\newcounter{hran}

\def\MSbar{\relax\ifmmode\overline{\rm MS}\else{$\overline{\rm MS}${ }}\fi}

\def\d{\rm d}

\def\vq{\vec{q}}

 \def\vx{\vec{ x}} 
\def\vk{\vec{k}}
\def\vy{\vec{y}}

\def\tr{{\rm tr}}

\numberwithin{equation}{section}
\begin{document}
\vspace{5mm}
\vspace{0.5cm}
\begin{center}

\def\thefootnote{\fnsymbol{footnote}}

{\large \bf 
  Operator Product Expansion   of Inflationary Correlators \\ [0.4cm]  and 
  Conformal Symmetry of de Sitter 
}
\\[1.5cm]
{\large  A. Kehagias$^{a}$ and A. Riotto$^{b}$}
\\[0.5cm]

\vspace{.3cm}
{\normalsize {\it  $^{a}$ Physics Division, National Technical University of Athens, \\15780 Zografou Campus, Athens, Greece}}\\

\vspace{.3cm}
{\normalsize { \it $^{b}$ Department of Theoretical Physics and Center for Astroparticle Physics (CAP)\\ 24 quai E. Ansermet, CH-1211 Geneva 4, Switzerland}}\\

\vspace{.3cm}


\end{center}

\vspace{3cm}

\hrule \vspace{0.3cm}
{\small  \noindent \textbf{Abstract} \\[0.3cm]
\noindent 
We study the multifield inflationary models where the cosmological perturbation  is sourced by light scalar fields other 
than the inflaton. The corresponding perturbations  are  both scale invariant and special conformally invariant. We exploit  the operator product expansion technique of conformal field theories to study the   inflationary correlators enjoying 
 the symmetries present during the de Sitter epoch. 
The operator product expansion  is particularly   powerful in  characterizing  inflationary correlation functions in two 
 observationally interesting limits, 
the squeezed limit of the three-point correlator and the collapsed limit of the four-point correlator.
Despite the fact that the shape of the four-point correlators is not fixed by the symmetries of de Sitter, its   
 exact shape can be found  in the collapsed limit making use of the operator product expansion.  By employing the fact that
 conformal invariance imposes  the two-point cross-correlations
 of the light fields to vanish unless the fields have the same conformal weights, we are able to show that the Suyama-Yamaguchi inequality relating the coefficients $f_{\rm NL}$ of the bispectrum in the squeezed limit and $\tau_{\rm NL}$ of the trispectrum in the collapsed limit 
also holds   when  the light fields are intrinsically non-Gaussian. In fact, we show that the inequality 
is valid irrespectively of the conformal symmetry, being just a consequence of fundamental 
 physical principles, such as 
the  short-distance expansion of operator products.
The observation of a strong  violation of the inequality will then have profound implications for inflationary models 
as it will imply either that  multifield inflation cannot be responsible for
generating the observed fluctuations independently of the details of the model or that some new non-trivial degrees of freedom play a role during inflation.

\vspace{0.5cm}  \hrule
\vskip 1cm

\def\thefootnote{\arabic{footnote}}
\setcounter{footnote}{0}



\baselineskip= 19pt

\newpage 
\tableofcontents

\newcommand{\fix}{\Phi(\mathbf{x})}
\newcommand{\fiLx}{\Phi_{\rm L}(\mathbf{x})}
\newcommand{\fiNLx}{\Phi_{\rm NL}(\mathbf{x})}
\newcommand{\fik}{\Phi(\mathbf{k})}
\newcommand{\fiLk}{\Phi_{\rm L}(\mathbf{k})}
\newcommand{\fiLkone}{\Phi_{\rm L}(\mathbf{k_1})}
\newcommand{\fiLktwo}{\Phi_{\rm L}(\mathbf{k_2})}
\newcommand{\fiLkthree}{\Phi_{\rm L}(\mathbf{k_3})}
\newcommand{\fiLkfour}{\Phi_{\rm L}(\mathbf{k_4})}
\newcommand{\fiNLk}{\Phi_{\rm NL}(\mathbf{k})}
\newcommand{\fiNLkone}{\Phi_{\rm NL}(\mathbf{k_1})}
\newcommand{\fiNLktwo}{\Phi_{\rm NL}(\mathbf{k_2})}
\newcommand{\fiNLkthree}{\Phi_{\rm NL}(\mathbf{k_3})}

\newcommand{\kernel}{f_{\rm NL} (\mathbf{k_1},\mathbf{k_2},\mathbf{k_3})}
\newcommand{\dirac}{\delta^{(3)}\,(\mathbf{k_1+k_2-k})}
\newcommand{\dirackonektwokthree}{\delta^{(3)}\,(\mathbf{k_1+k_2+k_3})}

\newcommand{\beq}{\begin{equation}}
\newcommand{\eeq}{\end{equation}}
\newcommand{\beqarr}{\begin{eqnarray}}
\newcommand{\eeqarr}{\end{eqnarray}}

\newcommand{\angk}{\hat{k}}
\newcommand{\angn}{\hat{n}}

\newcommand{\tfnow}{\Delta_\ell(k,\eta_0)}
\newcommand{\tf}{\Delta_\ell(k)}
\newcommand{\tfone}{\Delta_{\el\ell_1}(k_1)}
\newcommand{\tftwo}{\Delta_{\el\ell_2}(k_2)}
\newcommand{\tfthree}{\Delta_{\el\ell_3}(k_3)}
\newcommand{\tffour}{\Delta_{\el\ell_1^\prime}(k)}
\newcommand{\deltatilde}{\widetilde{\Delta}_{\el\ell_3}(k_3)}

\newcommand{\alm}{a_{\ell m}}
\newcommand{\almL}{a_{\ell m}^{\rm L}}
\newcommand{\almNL}{a_{\ell m}^{\rm NL}}
\newcommand{\almone}{a_{\el\ell_1 m_1}}
\newcommand{\almLone}{a_{\el\ell_1 m_1}^{\rm L}}
\newcommand{\almNLone}{a_{\el\ell_1 m_1}^{\rm NL}}
\newcommand{\almtwo}{a_{\el\ell_2 m_2}}
\newcommand{\almLtwo}{a_{\el\ell_2 m_2}^{\rm L}}
\newcommand{\almNLtwo}{a_{\el\ell_2 m_2}^{\rm NL}}
\newcommand{\almthree}{a_{\el\ell_3 m_3}}
\newcommand{\almLthree}{a_{\el\ell_3 m_3}^{\rm L}}
\newcommand{\almNLthree}{a_{\el\ell_3 m_3}^{\rm NL}}

\newcommand{\YLMstar}{Y_{L M}^*}
\newcommand{\Ylmstar}{Y_{\ell m}^*}
\newcommand{\Ylmstarone}{Y_{\el\ell_1 m_1}^*}
\newcommand{\Ylmstartwo}{Y_{\el\ell_2 m_2}^*}
\newcommand{\Ylmstarthree}{Y_{\el\ell_3 m_3}^*}
\newcommand{\Ylmstarfour}{Y_{\el\ell_1^\prime m_1^\prime}^*}
\newcommand{\Ylmstarfive}{Y_{\el\ell_2^\prime m_2^\prime}^*}
\newcommand{\Ylmstarsix}{Y_{\el\ell_3^\prime m_3^\prime}^*}

\newcommand{\YLM}{Y_{L M}}
\newcommand{\Ylm}{Y_{\ell m}}
\newcommand{\Ylmone}{Y_{\el\ell_1 m_1}}
\newcommand{\Ylmtwo}{Y_{\el\ell_2 m_2}}
\newcommand{\Ylmthree}{Y_{\el\ell_3 m_3}}
\newcommand{\Ylmfour}{Y_{\el\ell_1^\prime m_1^\prime}}
\newcommand{\Ylmfive}{Y_{\el\ell_2^\prime m_2^\prime}}
\newcommand{\Ylmsix}{Y_{\el\ell_3^\prime m_3^\prime}}

\newcommand{\comm}[1]{\textbf{\textcolor{rossos}{#1}}}
\newcommand{\lsim}{\,\raisebox{-.1ex}{$_{\textstyle <}\atop^{\textstyle\sim}$}\,}
\newcommand{\gsim}{\,\raisebox{-.3ex}{$_{\textstyle >}\atop^{\textstyle\sim}$}\,}

\newcommand{\jl}{j_\ell(k r)}
\newcommand{\jlfourone}{j_{\el\ell_1^\prime}(k_1 r)}
\newcommand{\jlfivetwo}{j_{\el\ell_2^\prime}(k_2 r)}
\newcommand{\jlsixthree}{j_{\el\ell_3^\prime}(k_3 r)}
\newcommand{\jlsix}{j_{\el\ell_3^\prime}(k r)}
\newcommand{\jlthree}{j_{\el\ell_3}(k_3 r)}
\newcommand{\jlthreetau}{j_{\el\ell_3}(k r)}

\newcommand{\Gaunt}{\mathcal{G}_{\el\ell_1^\prime \, \el\ell_2^\prime \, 
\el\ell_3^\prime}^{m_1^\prime m_2^\prime m_3^\prime}}
\newcommand{\Gaunttwo}{\mathcal{G}_{\el\ell_1^\prime \, \el\ell_2^\prime \, 
\el\ell_3}^{m_1^\prime m_2^\prime m_3}}
\newcommand{\Gauntstardef}{\mathcal{H}_{\el\ell_1 \, \el\ell_2 \, \el\ell_3}^{m_1 m_2 m_3}}
\newcommand{\Gauntstarone}{\mathcal{G}_{\el\ell_1 \, L \,\, \el\ell_1^\prime}
^{-m_1 M m_1^\prime}}
\newcommand{\Gauntstartwo}{\mathcal{G}_{\el\ell_2^\prime \, \el\ell_2 \, L}
^{-m_2^\prime m_2 M}}

\newcommand{\de}{{\rm d}}

\newcommand{\dangn}{d \angn}
\newcommand{\dangk}{d \angk}
\newcommand{\dangkone}{d \angk_1}
\newcommand{\dangktwo}{d \angk_2}
\newcommand{\dangkthree}{d \angk_3}
\newcommand{\dk}{d^3 k}
\newcommand{\dkone}{d^3 k_1}
\newcommand{\dktwo}{d^3 k_2}
\newcommand{\dkthree}{d^3 k_3}
\newcommand{\dkfour}{d^3 k_4}
\newcommand{\dallk}{\dkone \dktwo \dk}

\newcommand{\FT}{ \int  \! \frac{d^3k}{(2\pi)^3} 
e^{i\mathbf{k} \cdot \angn \eta_0}}
\newcommand{\planewave}{e^{i\mathbf{k \cdot x}}}
\newcommand{\dallkfourier}{\frac{\dkone}{(2\pi)^3}\frac{\dktwo}{(2\pi)^3}
\frac{\dkthree}{(2\pi)^3}}

\newcommand{\Bis}{B_{\el\ell_1 \el\ell_2 \el\ell_3}^{m_1 m_2 m_3}}
\newcommand{\Avbis}{B_{\el\ell_1 \el\ell_2 \el\ell_3}}

\newcommand{\los}{\mathcal{L}_{\el\ell_3 \el\ell_1 \el\ell_2}^{L \, 
\el\ell_1^\prime \el\ell_2^\prime}(r)}
\newcommand{\loszero}{\mathcal{L}_{\el\ell_3 \el\ell_1 \el\ell_2}^{0 \, 
\el\ell_1^\prime \el\ell_2^\prime}(r)}
\newcommand{\losone}{\mathcal{L}_{\el\ell_3 \el\ell_1 \el\ell_2}^{1 \, 
\el\ell_1^\prime \el\ell_2^\prime}(r)}
\newcommand{\lostwo}{\mathcal{L}_{\el\ell_3 \el\ell_1 \el\ell_2}^{2 \, 
\el\ell_1^\prime \el\ell_2^\prime}(r)}
\newcommand{\losfNL}{\mathcal{L}_{\el\ell_3 \el\ell_1 \el\ell_2}^{0 \, 
\el\ell_1 \el\ell_2}(r)}



\def\d{d}
\def\C{{\rm CDM}}
\def\me{m_e}
\def\te{T_e}
\def\ti{\tau_{\rm initial}}
\def\tci#1{n_e(#1) \sigma_T a(#1)}
\def\tr{\eta_r}
\def\dtr{\delta\eta_r}
\def\dd{\widetilde\Delta^{\rm Doppler}}
\def\dsw{\Delta^{\rm Sachs-Wolfe}}
\def\clsw{C_\ell^{\rm Sachs-Wolfe}}
\def\cldop{C_\ell^{\rm Doppler}}
\def\Dt{\widetilde{\Delta}}
\def\mut{\mu}
\def\vt{\widetilde v}
\def\hp{ {\bf \hat p}}
\def\sdv{S_{\delta v}}
\def\svv{S_{vv}}
\def\bvt{\widetilde{\bv}}
\def\delt{\widetilde{\delta_e}}
\def\cos{{\rm cos}}
\def\nn{\nonumber \\}
\def\bq{ {\bf q} }
\def\ba{ {\bf p} }
\def\bap{ {\bf p'} }
\def\bqp{ {\bf q'} }
\def\bp{ {\bf p} }
\def\bpp{ {\bf p'} }
\def\bk{ {\bf k} }
\def\bx{ {\bf x} }
\def\bv{ {\bf v} }
\def\qp{ p^{\mu}k_{\mu} }
\def\qpp { p^{\mu} k'_{\mu} }
\def\bgm{ {\bf \gamma} }
\def\bkp{ {\bf k'} }
\def\gq{ g(\bq)}
\def\gqp{ g(\bqp)}
\def\fp{ f(\bp)}
\def\h#1{ {\bf \hat #1}}
\def\fpp{ f(\bpp)}
\def\fz{f^{(\vec{0})}(p)}
\def\fpz{f^{(\vec{0})}(p')}
\def\f#1{f^{(#1)}(\bp)}
\def\fps#1{f^{(#1)}(\bpp)}
\def\dq{ {d^3\bq \over (2\pi)^32E(\bq)} }
\def\dqp{ {d^3\bqp \over (2\pi)^32E(\bqp)} }
\def\dpp{ {d^3\bpp \over (2\pi)^32E(\bpp)} }
\def\dtq{ {d^3\bq \over (2\pi)^3} }
\def\dtqp{ {d^3\bqp \over (2\pi)^3} }
\def\dtpp{ {d^3\bpp \over (2\pi)^3} }
\def\part#1;#2 {\partial#1 \over \partial#2}
\def\deriv#1;#2 {d#1 \over d#2}
\def\Done{\Delta^{(1)}}
\def\Dtwo{\widetilde\Delta^{(2)}}
\def\fone{f^{(1)}}
\def\ftwo{f^{(2)}}
\def\tg{T_\gamma}
\def\delpp{\delta(p-p')}
\def\delb{\delta_B}
\def\tc{\eta_0}
\def\DD{\langle|\Delta(k,\mu,\eta_0)|^2\rangle}
\def\DDL{\langle|\Delta(k=l/\tc,\mu)|^2\rangle}
\def\bkpp{{\bf k''}}
\def\kmkp{|\bk-\bkp|}
\def\kmkpsq{k^2+k'^2-2kk'x}
\def\tt{ \left({\tau' \over \tau_c}\right)}
\def\kt{ k\mu \tau_c}

%
%
%

\section{Introduction}
\noindent
One of the basic ideas of modern cosmology is that there was an epoch early
in the history of the universe when potential, or vacuum, energy 
associated to a scalar field, the inflaton, 
dominated other forms of energy density such as matter or radiation. 
During such a
vacuum-dominated era the scale factor grew exponentially (or nearly
exponentially) in time. During this phase, dubbed inflation 
\cite{guth81,lrreview},
a small,  smooth spatial region of size less than the Hubble radius
could grow so large as to easily encompass the comoving volume of the 
entire presently observable universe. If the universe underwent
such a period of rapid expansion, one can understand why the observed
universe is so homogeneous and isotropic to high accuracy.

Inflation has also become the dominant 
paradigm for understanding the 
initial conditions for structure formation and for Cosmic
Microwave Background (CMB) anisotropy. In the
inflationary picture, primordial density and gravity-wave fluctuations are
created from quantum fluctuations ``redshifted'' out of the horizon during an
early period of superluminal expansion of the universe, where they
are ``frozen'' \cite{muk81,hawking82,starobinsky82,guth82,bardeen83}. 
Perturbations at the surface of last scattering are observable as temperature 
anisotropy in the CMB. 
The last and most impressive confirmation of the inflationary paradigm has 
been recently provided by the data 
of the Wilkinson Microwave Anistropy Probe (WMAP) mission which has 
marked the beginning of the precision era of the CMB measurements in space
\cite{wmap7}.
The WMAP collaboration has  produced a full-sky map of the angular variations 
of the CMB, with unprecedented accuracy.
WMAP data confirm the inflationary mechanism as responsible for the
generation of curvature (adiabatic) superhorizon fluctuations. 

Despite the simplicity of the inflationary paradigm, the mechanism
by which  cosmological adiabatic perturbations are generated  is not
yet fully established. In the standard picture, the observed density 
perturbations are due to fluctuations of the inflaton field itself. 
When inflation ends, the inflaton oscillates about the minimum of its
potential and decays, thereby reheating the universe. As a result of the 
fluctuations
each region of the universe goes through the same history but at slightly
different times. The 
final temperature anisotropies are caused by the fact that
inflation lasts different amounts of time in different regions of the universe
leading to adiabatic perturbations. Under this hypothesis, 
the WMAP dataset already allows
to extract the parameters relevant 
for distinguishing among single-field inflation models.

An alternative to the standard scenario is represented by the curvaton 
mechanism
\cite{curvaton1,LW,curvaton3} where the final curvature perturbations
are produced from an initial isocurvature perturbation associated to the
quantum fluctuations of a light scalar field (other than the inflaton), 
the curvaton, whose energy density is negligible during inflation. The 
curvaton isocurvature perturbations are transformed into adiabatic
ones when the curvaton decays into radiation much after the end 
of inflation. Alternatives to the curvaton model are
those models characterized by the curvature perturbation being generated by an inhomogeneity in
the decay rate \cite{rate} or the mass \cite{mass} of the particles responsible for the reheating after inflation.
Other opportunities for generating the curvature perturbation occur at the end of inflation \cite{end} and
during preheating \cite{during}. 

All these alternative models to generate the cosmological perturbations have in common that the comoving curvature perturbation in generated on superhorizon scale when the isocurvature perturbation, which is  associated to the fluctuations of these light scalar fields different from the inflaton,  is converted
into curvature perturbation after (or at the end) of inflation. 
The very simple fact that during inflation the fluctuation associated to these light fields  is of the isocurvature type, that is the energy density stored in these fields is small compared to the vacuum energy responsible for inflation, implies that the de Sitter isometries are not broken by the presence of these light fields. Therefore their statistical correlators should enjoy all the symmetries present during  the de Sitter epoch and therefore be not only scale invariant, but also conformal invariant. Building up on the results of  Ref. \cite{mp} (where the most general three-point function for gravitational waves produced during a period of exactly de Sitter expansion was studied)  and of  
Ref. \cite{anto},  in Ref. \cite{Creminelli1}  the consequences of scale invariance and special conformal symmetry of scalar perturbations were discussed. Further extensions appeared in Ref.  
 \cite{Creminelli2} where conformal consistency relations for single-field inflation have been investigated  and in Ref. \cite{new} where the existence of non-linearly realized conformal symmetries for scalar adiabatic perturbations in cosmology has been pointed out.

In this paper we are concerned with the large class of multifield models where the non-Gaussianity (NG) of the curvature perturbation is sourced by light fields other than the inflaton.
By the $\delta N$ formalism \cite{deltaN}, the comoving curvature perturbation $\zeta$ 
on a uniform
energy density hypersurface at time $t_{\rm f}$ is, on sufficiently large scales, equal to the perturbation
in the time integral of the local expansion from an initial flat hypersurface ($t = t_{*}$)
to the final uniform energy density hypersurface. On sufficiently large scales, the local expansion
can be approximated quite well by the expansion of the unperturbed Friedmann
universe. Hence the curvature perturbation at time $t_{\rm f}$ can be expressed in terms of  the values of the relevant scalar
fields $\sigma^I(t_{*},\vec{x})$ at $t_{*}$

\be
\zeta(t_{\rm f},\vec{x})=N_I\sigma^I+\frac{1}{2}N_{IJ}\sigma^I\sigma^J+\cdots, \label{zeta}
\label{deltan}
\ee
where $N_I$ and $N_{IJ}$ are the first and second derivative, respectively, of the number of e-folds 
\be
N(t_{\rm f},t_{*},\vec{x})=\int_{t_{*}}^{t_{\rm f}}\,{\rm d}t\, H(t,\vec{x}).
\ee
with respect to the 
field $\sigma^I$. From the expansion (\ref{deltan}) one can read off the $n$-point correlators. 
For instance, the three- and four-point correlators of the comoving curvature perturbation, the so-called bispectrum and trispectrum respectively,  is given by
\be
B_\zeta(\vk_1,\vk_2,\vk_3)=
N_I N_J N_K B^{IJK}_{\vec{k}_1\vec{k}_2\vec{k}_3}+ N_I N_{JK}N_{L}\left(P^{IK}_{\vec{k}_1}P^{JL}_{\vec{k}_2}+2\,\,{\rm permutations} \label{zeta3}
\right)
\ee
and
\begin{eqnarray}
T_\zeta(\vk_1,\vk_2,\vk_3,\vk_4)&=&
N_I N_J N_K N_L T^{IJKL}_{\vec{k}_1\vec{k}_2\vec{k}_3\vec{k}_4}\nonumber\\
&+& N_{IJ} N_{K}N_{L}N_M\left(P^{IK}_{\vec{k}_1}B^{JLM}_{\vec{k}_{12}\vec{k}_3\vec{k}_4}+11\,\,{\rm permutations}
\right)\nonumber\\
&+&N_{IJ} N_{KL}N_{M}N_N\left(P^{JL}_{\vec{k}_{12}}P^{IM}_{\vec{k}_{1}}P^{KN}_{\vec{k}_{3}}
+11\,\,{\rm permutations}
\right)
\nonumber\\
&+&N_{IJK} N_{L}N_{M}N_N\left(P^{IL}_{\vec{k}_{1}}P^{JM}_{\vec{k}_{2}}P^{KN}_{\vec{k}_{3}}
+3\,\,{\rm permutations}
\right),
 \label{zeta4}
\end{eqnarray}
where 

\begin{eqnarray}
\langle\sigma_{\vec{k}_{1}}^I\sigma^J_{\vec{k}_{2}}\rangle&=&(2\pi)^3\delta({\vec{k}_{1}}+{\vec{k}_{2}})P^{IJ}_{\vec{k}_{1}},\nonumber\\
\langle\sigma_{\vec{k}_{1}}^I\sigma^J_{\vec{k}_{2}}\sigma^K_{\vec{k}_{3}}\rangle&=&(2\pi)^3\delta({\vec{k}_{1}}+{\vec{k}_{2}}+{\vec{k}_{3}})B^{IJK}_{\vec{k}_{1}\vec{k}_{2}\vec{k}_{3}},\nonumber\\
\langle\sigma_{\vec{k}_{1}}^I\sigma^J_{\vec{k}_{2}}\sigma^J_{\vec{k}_{3}}\sigma^L_{\vec{k}_{4}}\rangle&=&(2\pi)^3\delta({\vec{k}_{1}}+{\vec{k}_{2}}+{\vec{k}_{3}}{\vec{k}_{4}})T^{IJKL}_{\vec{k}_{1}\vec{k}_{2}\vec{k}_{3}\vec{k}_{4}}, 
\end{eqnarray}
 and $\vec{k}_{ij}=(\vec{k}_{i}+\vec{k}_{j})$. We see that  
the three-point correlator (and similarly for the four-point one) of the comoving curvature perturbation is  the sum of two pieces. 
One, proportional to the three-point correlator of the $\sigma^I$ fields,  is model-dependent and present when the   fields $\sigma^I$ are intrinsically NG. The second one is universal and   is generated when the modes of 
the fluctuations are superhorizon and is present even if the $\sigma^I$ fields are gaussian. 
One should keep in mind that the relative magnitude between the two contributions is 
model-dependent 
and that the constraints imposed by the symmetries present during the de Sitter stage apply 
separately to both the first and the second contribution\footnote{
The reason is that although the scalar fields $\sigma^I$ may have specific scaling dimension and may
 transform irreducibly 
under the conformal group, the comoving curvature perturbations  $\zeta$ does not have specific scaling dimension as it is the sum of 
operators with different dimensions. In other words,  $\zeta$ is a reducible representation of the conformal group. However, its 
$n$-point functions may be specified by the conformal properties of its irreducible components of the conformal group.}. 
Even though the intrinsically NG contributions to the $n$-point correlators are model-dependent, 
their forms are dictated
by the conformal symmetry of the de Sitter stage (although their amplitudes remain
model-dependent). This is the subject of the present paper. 

After a brief summary in  of the symmetries of the de Sitter geometry in  section 2, we will discuss in section 3 the constraints imposed by scale invariance and conformal symmetry on the two- and three-point correlators. In particular, we will demonstrate that the two-point cross-correlations of the light fields vanish unless their conformal weights are equal. This is a in fact a standard result of field theories enjoying conformal symmetry. 

We then turn out attention to the 
operator product expansion technique of conformal field theories to investigate which kind of   informations we can gather on inflationary correlations  for fields considered at  coincidence points. The operator product expansion is very  powerful to analyze the squeezed limit of the bispectrum and the collapsed limit of the trispectrum. 
These limits 
 are  particularly interesting from the
 observationally point of view because they are associated to the local model of NG  
 (for a review see  \cite{revNG}) which  leads to pronounced effects of NG on the clustering of dark matter halos and to  strongly scale-dependent bias \cite{dalal}. 
 
 We use the techniques developed by  Ferrara, Gatto and Grillo in the early 70's to find the model-independent shape of the three- and point-correlators in the squeezed and collapsed limit, respectively. While conformal symmetry does not fix uniquely the shape of the four-point correlator, we show that its shape can be indeed computed in the collapsed limit by using the so-called conformal blocks. This allows us to prove that  the contribution  to the three- and four-point correlators of the curvature perturbation 
 from the connected three- and four-point correlators of the $\sigma^I$ fields (originated from the fact that these fields can be  intrinsically NG) have the same shapes of the     universal and model-independent contribution generated when the modes of 
the fluctuations are superhorizon and  present even if the $\sigma^I$ fields are gaussian. This is done in section 4.  
This result allows us to extend in section 5 the so-called 
Suyama-Yamaguchi  inequality \cite{SY} which relates the coefficient of the trispectrum  $\tau_{\rm NL}$ of the curvature perturbation 
in the collapsed limit to the coefficient $f_{\rm NL}$ of the squeezed limit of its bispectrum 
and was proved under the condition that   the fluctuations of the  scalar fields $\sigma^I$ at 
the horizon crossing are scale invariant and gaussian. 
A generalization of this inequality was provided in Refs. \cite{SY1} to the case of NG $\sigma^I$ fields. However there the  crucial assumption was made that   the coefficients $f_{\rm NL}$ and $\tau_{\rm NL}$ were momentum-independent, which is not automatically guaranteed if the fields are NG. Based on our results stemming from scale invariance and special conformal symmetry, we can show that indeed $f_{\rm NL}$ and $\tau_{\rm NL}$ are momentum-independent
in the squeezed and collapsed limits respectively and therefore 
we are able to show that the Suyama-Yamaguchi   inequality is  valid when the light 
fields $\sigma^I$ are NG.    In fact, we will  take a  further step and,  based on the operators' 
short-distance expansion, we will prove that  that the Suyama-Yamaguchi inequality holds 
on general grounds. It is consequence of fundamental physical principles (like positivity of the two-point function) and its hard violation 
would required some new non-trivial physics to be involved.
The observation of a strong  violation of the inequality will then have profound implications for inflationary models 
as it will imply either that  multifield inflation cannot be responsible for
generating the observed fluctuations independently of the details of the model or that some new non-trivial degrees of freedom play a role during inflation.

In section 6 we study, even though briefly,  the possible implications of another  class of conformal  theories, namely the logarithmic conformal field theories,   which can be of interest from the cosmological point of view. 
These are theories characterized by the appearance
of logarithms in correlation functions  due to logarithmic short-distance singularities in the operator product expansion. As a consequence,  the spectral index of the curvature perturbation power spectrum  gets a new contribution  
 due to logarithmic short-distance singularities in the OPE. This contribution is present  even if the fields light involved are massless. 
Finally, section 7 present our conclusions.

\section{Symmetries of the de Sitter geometry}
\noindent
Conformal invariance 
in three-dimensional space $\mathbb{R}^3$ is connected to the symmetry under the group $SO(1,4)$ in the same way 
conformal invariance in a four-dimensional Minkowski spacetime is connected to the $SO(2,4)$ group. As $SO(1,4)$ is the 
isometry group of de Sitter spacetime,  a conformal phase during which fluctuations were generated could be 
a  de Sitter stage.  In such a  case, the kinematics  is specified by the embedding 
of $\mathbb{R}^3$ as flat sections in de Sitter spacetime. The de Sitter isometry group acts  as conformal group 
on $\mathbb{R}^3$ when the fluctuations are superhorizon. It is in this regime that  the $SO(1,4)$ isometry
of the de Sitter background is realized as conformal symmetry of the flat $\mathbb{R}^3$ sections \cite{anto,Creminelli1}.  Correlators are expected to be constrained by conformal invariance. All these reasonings apply in the case in which the cosmological perturbations are generated 
by light scalar fields other than the inflaton. Indeed, it is only in such a case that correlators
inherit all the isometries of de Sitter. 

It is also important to stress that  the 
two-point correlator cannot capture the  full conformal invariance and  is 
only sensitive to the scale invariance symmetry.  To reveal the full conformal symmetry 
one needs to consider higher-order correlators. This is what we will do in the following.
 Before though,  and  for the sake of self-completeness,  we would like to remind the
reader  of some basic geometrical and algebraic properties of de Sitter spacetime and group \cite{HE}. The expert reader may skip the following two sections many details of which are contained already in, for instance,  Ref. \cite{anto}.

The four-dimensional de Sitter spacetime of radius $H^{-1}$ is described by the hyperboloid 
\be
\eta_{AB}X^AX^B=-X_0^2+X_i^2+X_5^2=\frac{1}{H^2} ~~~~( i=1,2,3),  \label{hyper}
\ee
embedded in  five-dimensional Minkowski spacetime $\mathbb{M}^{1,4}$  with coordinates $X^A$ 
and flat metric  $\eta_{AB}=\rm{diag}(-1,1,1,1,1)$.
A particular parametrization of the de Sitter hyperboloid is provided by
\be
&&X^0=\frac{1}{2H}\left(H\eta-\frac{1}{H \eta}\right)-\frac{1}{2}\frac{x^2}{\eta},\nonumber \\
&&X^i=\frac{x^i}{H\eta},\nonumber \\
&&X^5=-\frac{1}{2H}\left(H\eta+\frac{1}{H \eta}\right)+\frac{1}{2}\frac{x^2}{\eta},  \label{poin}
\ee
which may easily be checked that satisfies Eq. (\ref{hyper}). The de Sitter metric is the induced metric on the hyperboloid from 
the five-dimensional ambient Minkowski spacetime
\be
{\rm d} s_5^2=\eta_{AB}dX^AdX^B.
\ee
For the particular parametrization (\ref{poin}), for example,  we find
\be
 {\rm d}s^2=\frac{1}{H^2\eta^2}\left(-{\rm d}\eta^2+{\rm d} \vx^2\right).
\ee 
The group $SO(1,4)$ acts linearly on $\mathbb{M}^{1,4}$. Its generators are
\be
J_{AB}=X_A\frac{\partial}{\partial X^B}-X_B\frac{\partial}{\partial X^A}\, ~~~A,B=(0,1,2,3,5)
\ee
and  satisfy the $SO(1,4)$ algebra
\be
[J_{AB},J_{CD}]=\eta_{AD} J_{BC}-\eta_{AC}J_{BD}+\eta_{BC}J_{AD}-\eta_{BD}J_{AC}.
\ee
We may split these generators as 
\be 
\label{JP+}
J_{ij},~~ P_0=J_{05}\, , ~~\Pi^+_i=J_{i5}+J_{0i}\, , ~~\Pi^-_i=J_{i5}-J_{0i},
\ee
which act on the de Sitter hyperboloid as 
\be
&&J_{ij}=x_i\frac{\partial}{\partial x_j}-x_j\frac{\partial}{\partial x_i},\nonumber \\
&&P_0=\eta\frac{\partial}{\partial\eta}+x^i\frac{\partial}{\partial x^i},\nonumber \\
&& \Pi^-_i=-2H \eta x^i\frac{\partial}{\partial\eta}+H\left(x^2\delta_{ij}-2 x_i x_j\right)\frac{\partial}{\partial x_j}
-H \eta^2 \frac{\partial}{\partial x_i},\nonumber \\
&&\Pi^+_i=\frac{1}{H}\frac{\partial}{\partial x_i}
\ee
and satisfy the commutator relations
\be
\label{dsit}
&&[J_{ij},J_{kl}]=\delta_{il} J_{jk}-\delta_{ik}J_{jl}+\delta_{jk}J_{il}-\delta_{jl}J_{ik},\nonumber \\
&&[J_{ij},\Pi^{\pm}_k]=\delta_{ik} \Pi^{\pm}_{j}-\delta_{jk}\Pi^{\pm}_i,\nonumber\\
&&[\Pi^{\pm}_k,P_0]=\mp \Pi^{\pm}_k, \nonumber\\
&&[\Pi^{-}_i,\Pi^{+}_j]=2J_{ij}+2\delta_{ij}P_0.
\ee
This is nothing else that the conformal algebra. Indeed, by defining 
\be
\label{LP}
L_{ij}=iJ_{ij}\, , ~~~D=-i P_0\, , ~~~P_i=-i\Pi^+_i\, , ~~~K_i=i\Pi^-_i, 
\ee
we get 
\be
&&P_i=-\frac{i}{H}\partial_i, \nonumber\\ 
&&D=-i\left(\eta\frac{\partial}{\partial\eta}+x^i\partial_i\right), \nonumber \\
&&K_i=-2iHx_i\left(\eta\frac{\partial}{\partial\eta}+x^i\partial_i\right)-iH(-\eta^2+x^2)\partial_i, \nonumber\\
&&L_{ij}=i\left(x_i\frac{\partial}{\partial x_j}-x_j\frac{\partial}{\partial x_i}\right). \label{jpkd}
\ee
These are also the Killing vectors of de Sitter spacetime corresponding to symmetries under space translations ($P_i$), dilitations 
($D$), special conformal transformations ($K_i$) and space rotations ($L_{ij}$).  
They satisfy
the conformal algebra in its standard form 
\be
\label{conf}
&&[D,P_i]=i P_i, \\
&&[D,K_i]=-iK_i \label{DK}, \\
&&[K_i,P_j]=2i\Big{(}\delta_{ij}D-L_{ij}\Big{)} \\
&&[L_{ij},P_k]=i\Big{(}\delta_{jk}P_i-\delta_{ik}P_j\Big{)}, \\
&&[L_{ij},K_k]=i\Big{(}\delta_{jk}K_i-\delta_{ik}K_j\Big{)}, \\
&&[L_{ij},D]=0, \label{LD}\\
&&[L_{ij},L_{kl}]=i\Big{(}\delta_{il} L_{jk}-\delta_{ik}L_{jl}+\delta_{jk}L_{il}-\delta_{jl}L_{ik}\Big{)}.
\ee
The de Sitter algebra $SO(1,4)$ has two Casimir invariants
\be
&&{\cal{C}}_1=-\frac{1}{2}J_{AB}J^{AB}\, , ~~~\\
&&{\cal{C}}_2 = W_A W^A\, , ~~~~~W^A=\epsilon^{ABCDE}J_{BC} J_{DE}.
\ee
Using Eqs. (\ref{JP+}) and (\ref{LP}), we find that 
\be
{\cal{C}}_1=D^2+\frac{1}{2}\{P_i,K_i\}+\frac{1}{2}L_{ij}L^{ij} \label{c1}, 
\ee
which turns out to be, in  the explicit  representation  Eq. (\ref{jpkd}),
\be
H^{-2}{\cal{C}}_1=-\frac{\partial^2}{\partial\eta^2}-\frac{2}{\eta}\frac{\partial}{\partial\eta}+\nabla^2.
\ee
As a result, ${\cal{C}}_1$ is the Laplace operator on the de Sitter hyperboloid and for a  scalar field $\phi(x)$ 
we have 
\be
{\cal{C}}_1\phi(x)=\frac{m^2}{H^2} \phi(x).
\ee
Let us now consider the case $H\eta\ll 1$.  The parametrization (\ref{poin}) turns out then to be 
\be
&&X^0=-\frac{1}{2H^2\eta}-\frac{1}{2}\frac{x^2}{\eta},\nonumber \\
&&X^i=\frac{x^i}{H\eta}\nonumber, \\
&&X^5=-\frac{1}{2H^2\eta}+\frac{1}{2}\frac{x^2}{\eta}
\ee
and we may easily check that the hyperboloid has been degenerated to the hypercone
\be
-X_0^2+X_i^2+X_5^2=0  \label{cone}.
\ee
We identify points $X^A\equiv \lambda X^A$ (which turns the cone (\ref{cone}) into a projective space).
 As a result, $\eta$ in the denominator
of the $X^A$ can be ignored due to projectivity condition. 
Then, on the cone,  the conformal group  acts linearly, whereas induces the (non-linear) conformal transformations $x_i\to x_i'$ with 
\be
&& x_i'=a_i+M_i^jx_j, \\
&& x_i'=\lambda x_i, \\
&&x_i'=\frac{x_i+b_ix^2}{1+2b_ix_i+b^2x^2}. \label{specconf}
\ee 
on Euclidean $\mathbb{R}^3$ with coordinates $x^i$.  These transformations correspond to translations and rotations (generated by $P_i,L_{ij}$), dilations
(generated by $D$) and special conformal transformations (generated by $K_i$), respectively, acting now on the constant time hypersurfaces  
of de Sitter spacetime.   It should be noted that special conformal transformations can be written in terms of inversion
\be
x_i\to x_i'=\frac{x_i}{x^2} \label{inv}
\ee
as  inversion$\times$translation$\times$inversion.

\subsection{Representations}
\noindent
The representations of the $SO(1,4)$ algebra are constructed by employing the method 
of  induced representations. Let us consider 
the stability subgroup at $x^i=0$ which is the group $G$ generated by  $(L_{ij},D,K_i)$. 
It is easy to see from the conformal algebra, that $P_i$ and $K_i$ are actually raising and lowering operators for the dilation
operator $D$. Therefore there should be  states which will be annihilated by $K_i$. Every irreducible representation  will then  
be specified by an irreducible representation of the rotational group $SO(3)$ ({\it i.e.} its spin) and a definite conformal dimension 
annihilated by $K_i$. Representations  $\phi_s(\vec{0})$ of the stability group at $\vx=\vec 0$ 
with spin $s$ and dimension $\Delta$   are  specified by    
\be
&&[L_{ij},\phi_s(\vec{0})]=\Sigma^{(s)}_{ij}\phi_s(\vec{0}), \nonumber \\
&& [D,\phi_s(\vec{0})]=-i \Delta \phi_s(\vec{0}), \nonumber \\
&&[K_i,\phi_s(\vec{0})]=0,  \label{ls}
\ee
where $\Sigma^{(s)}_{ij}$ is a spin-$s$ representation of $SO(3)$. 
Those representations $\phi_s(\vec{0})$  that satisfy the relations (\ref{ls}) are called primary fields. Once the primary fields are known, all other fields, 
the descendants,  are constructed by taking derivatives
of the primaries $\partial_i\cdots \partial_j \phi_s(\vec{0})$.   
For scalars in particular, the momentum  $P_i$ generates translations so that for a scalar $\phi(\vec{x})$
we have 
\be
[P_i,\phi(\vec{x})]=-i\partial_i \phi(\vec{x}).
\ee
Denoting collectively any generator of the stability subgroup $G$ as $J=(L_{ij},D,K_i)$ and 
taking into account that  $\phi(\vec{x})=e^{i\vec{P}\cdot \vx}\phi(\vec{0})e^{-i\vec{P}\cdot \vx}$, we find that 
\be
[J,\phi(\vec{x})]=e^{i\vec{P}\cdot \vx}[\hat{J},\phi(\vec{0})]e^{-i\vec{P}\cdot \vx}
\ee
where 
\be
\hat{J}=e^{-i\vec{P}\cdot \vx}Je^{i\vec{P}\cdot \vx}=\sum_n\frac{(-i)^n}{n!}x^{i_1}x^{i_2}\ldots x^{i_n}[P_{i_1}[P_{i_2}\ldots[P_{i_n},J],\ldots ]]
\ee
and $\phi(\vec{0})$ is a representation of the stability subgroup. 
Specifying for $J=L_{ij},D$ and and $J=K_i$ we find
\be
&&\hat{L}_{ij}=L_{ij}+x_iP_j-x_j P_i, \\
&&\hat{D}=D+x^i P_i,\\
&&\hat{K}_i=K_i+2(x_i D-x^jL_{ij})+2x_i x^jP_j-x^2P_i.
\ee
For a scalar, the right-hand side of the first equation  in (\ref{ls}) vanishes, 
therefore we find that for a scalar $\phi(\vec{x})$ 
\be
&&i [L_{ij},\phi(\vec{x})]=\left(x_i\partial_j-x_j\partial_i\right)\phi(\vec{x}), \\
&&i[K_i,\phi(\vec{x})]=\left(2\Delta x_i+2x_i x^j \partial_j-x^2\partial_i\right)\phi(\vec{x}), \\
&&i[D,\phi(\vec{x})]=\left(x^i\partial_i+\Delta\right)\phi(\vec{x}), \\
&&i[P_i,\phi(\vec{x})]=\partial_i \phi(\vec{x}).
\ee
Note that Eq. (\ref{c1}) gives, for example, 
\be
[{\cal{C}}_1,\phi(\vec{0})]=-\Delta(\Delta-3)\phi(\vec{0}), 
\ee
which implies that 
\be
m^2=-\Delta(\Delta-3)H^2. \label{MH}
\ee
 It can be shown that the scalar representations of the de Sitter group $SO(1,4)$ actually splits 
into three distinct series \cite{s1,s2,strominger}: the principal series with masses $m^2\geq 9H^2/4$, the complementary series with 
masses in the 
range $0<m^2<9H^2/4$ and the discrete series. It is the principal representations which survive the Winger-Inon\"u contraction
($H\to 0$) to the Poinc\'are group. 
  
The method of the induced representations used above for the scalar can  be employed to include higher-spin fields as well. For 
a higher-spin field described by a symmetric-traceless tensor $\phi_{i_1\ldots i_s}$ we get 
\be
&&i [L_{ij},\phi_{k_1\ldots k_s}]=\left(x_i\partial_j-x_j\partial_i+i\Sigma_{ij}^{(s)}\right)\phi_{k_1\ldots k_s}, \\
&&i[K_i,\phi_{k_1\ldots k_s}]=\left(2\Delta x_i+2x_i x^j \partial_j-x^2\partial_i+2i x^j\Sigma_{ji}^{(s)}\right)\phi_{k_1\ldots k_s}, \\
&&i[D,\phi_{k_1\ldots k_s}]=\left(x^i\partial_i+\Delta_s\right)\phi_{k_1\ldots k_s}, \\
&&i[P_i,\phi_{k_1\ldots k_s}]=\partial_i \phi_{k_1\ldots k_s},
\ee
where the spin operator $\Sigma_{ij}^{(s)}$ acts as 
\be
\Sigma_{ij}^{(s)}\phi_{k_1\ldots k_s}=\sum_{\{a\}} 
(\phi_{k_1\ldots k_{a-1}ik_{a+1}\ldots k_s}\delta_{jk_a}-\phi_{k_1\ldots k_{a-1}jk_{a+1}\ldots k_s}\delta_{ik_a}). \label{sigma}
\ee
It is then easy to verify that 
\be
{\cal{C}}_1=\frac{m^2}{H^2}=-\Delta_s(\Delta_s-3) -s(s+1)\,\,\,\,{\rm since }\,\, \,\,\frac{1}{2}\Sigma^{(s)}_{ij}\Sigma^{(s)}_{ij}=s(s+1).
\ee

\section{Symmetry Constraints}
\noindent
Let us consider now the constraints imposed by scale and conformal invariance to the $n$-point correlators. These constraints should be imagined to be applied to the light scalar fields $\sigma^I$ generating the comoving curvature perturbations after (or at the end) of inflation when the isocurvature modes they carry become the curvature mode.

\subsection{Scale Invariance}\label{sectionscale}
\noindent
Rotations, translations and dilations form a subgroup of the full conformal group. We would like first to explore the constraints this subgroup
imposes on the correlators. Obviously, rotation and translation invariance require correlators of the  operators at points 
$\vx_1$ and $ \vx_2$ to 
depend on $|\vx_1-\vx_2|$. As is well known,  the correlator of two operators
is completely 
determined by their scale dimensions whereas the functional form of three-point correlator is also determined by their dimensions.
Taking into account  also the scaling of operators under dilations, 
one finds that two- and three-point functions are specified to be
\be
&&\langle\sigma^I(\vx_1)\, \sigma^J(\vx_2)\rangle=\frac{c_{IJ}}{|\vx_1-\vx_2|^{\Delta_I+\Delta_J}},\label{2p} \\
&&\langle\sigma^I(\vx_1)\sigma^J(\vx_2)\sigma^K(\vx_3)\rangle=
\frac{c_{IJK}}{|\vx_1-\vx_2|^{w_K}|\vx_2-\vx_3|^{w_I}|\vx_3-\vx_1|^{w_J}},\label{3p}
\ee
where $c_{IJ}$ and $c_{IJK}$ are constants setting the amplitude of the correlators, 
$\sigma_{I,J,K}$ are operators of dimensions $\Delta_{I,J,K}$ and $
(w_I+w_J+w_K)=\Delta_I+\Delta_J+\Delta_K=3\Delta$. 
It is straightforward to write two- and three-point correlators in momentum space\footnote{The prime denotes correlators without the $(2\pi)^3\delta^{(3)}(\sum_i\vk_i)$ factors. 
} 

\be
\langle\sigma^I_{\vk_1}\sigma^J_{\vk_2}\rangle^\prime&=&c_{IJ}
\, k_1^{\Delta_I+\Delta_J-3}, 
\label{2pk}\\
\langle\sigma^I_{\vk_1}\sigma^J_{\vk_2}\sigma^K_{\vk_3}\rangle^\prime&=& 
c_{IJK}\prod_{i=I,J,K} \frac{2^{3-w_I}\pi^{3/2}\Gamma(\frac{3-w_I}{2})}
{\Gamma(\frac{w_I}{2})}
\nonumber \\
&\times& 
\int {\rm d}^3q \, |\vq|^{w_K-3}|\vq-\vk_1|^{w_J-3}|\vq+\vk_2|^{w_I-3}+\,{\rm cyclic}\nonumber \\
&=&
c_{IJK}2^{7-3\Delta}\pi^{\frac{5}{2}}\frac{\Gamma(3-\frac{3\Delta}{2})\Gamma(\frac{3-\Delta_K}{2})}{\Gamma(\frac{\Delta_I}{2})
\Gamma(\frac{\Delta_J}{2})}\nonumber \\
&\times&
k_1^{3\Delta-6}\int_0^1 {\rm d}u\frac{(1-u)^{\frac{1}{2}-\frac{\Delta_I}{2}}u^{\frac{1}{2}-\frac{\Delta_J}{2}}}{
\left[(1\!-\!u) X\!+\!u Y\right]^{3-\frac{3\Delta}{2}}} \, \,{}_2F_1\!\!\left(3\!-\!\frac{3\Delta}{2},\frac{\Delta_K}{2},
\frac{3}{2},{\cal{Z}}\right)+\,{\rm cyclic},
\label{3pk}
\ee
where we have used the definitions \cite{anto}
\be
X=\frac{k_2^2}{k_1^2}\, , ~~~~Y=\frac{k_3^2}{k_1^2}\, , ~~~~
{\cal{Z}}=1\!-\!\frac{u(1\!-\!u)}{
(1\!-\!u)X\!+\!uY}.
\ee
Eq. (\ref{3pk}) for the case $\Delta_{I}=\Delta_J=\Delta_K=w$ appeared in Ref. \cite{anto}. However, we note that the hypergeometric
function in (\ref{3pk}) converges in $|{\cal{Z}}|\leq1$ for 
\be
\Delta_I+\Delta_J> 3 \label{dd3}
\ee
and similarly for any pair of $\Delta$'s. Recalling that  the scaling dimensions are related to the masses as in Eq. (\ref{MH})),
\be
\Delta_{I,J,K}=\frac{3}{2}\left(1-\sqrt{1-\frac{4m_{I,J,K}^2}{9H^2}}\right),  
\ee
we see that  $0\leq \Delta_{I,J,K}\leq 3/2$ and Eq. (\ref{dd3}) is never satisfied. What we can do is to use Euler's transformation of
the hypergeometric function
\be
{}_2F_1(a,b,c,z)=(1-z)^{c-a-b}{}_2F_1(c-a,c-b,c,z)
\label{euler}
\ee
to express the three-point function in Eq. (\ref{3pk}) as 
\be
\langle\sigma^I_{\vk_1}\sigma^J_{\vk_2}\sigma^K_{\vk_3}\rangle^\prime
&=&c_{IJK}2^{7-3\Delta}\pi^{\frac{5}{2}}\frac{\Gamma(3-\frac{3\Delta}{2})\Gamma(\frac{3-\Delta_K}{2})}{\Gamma(\frac{\Delta_I}{2})
\Gamma(\frac{\Delta_J}{2})}\times\nonumber \\
&\times&
k_1^{3\Delta-6}\int_0^1 {\rm d}u\frac{(1-u)^{\frac{\Delta_J}{2}-1}u^{\frac{\Delta_I}{2}-1}}{
\left[(1\!-\!u) X\!+\!u Y\right]^{\frac{3-\Delta_K}{2}}} \, \,{}_2F_1\!\!\left(\frac{3\Delta}{2}\!-\!\frac{3}{2},\frac{3-\Delta_K}{2},
\frac{3}{2},{\cal{Z}}\right)+\,{\rm cyclic},
\label{3pkc}
\ee
which is now converging for  $0\leq \Delta_{I,J,K}\leq 3/2$. 
We stress that   the forms (\ref{2pk}) and (\ref{3pkc}) of the two- and three-point functions respectively, are dictated simply by 
scale invariance and not by special conformal symmetry.  
Full conformal invariance give additional constraints. 

We may also consider particular limits of the three-point function. As we wrote in the introduction, the so-called squeezed limit
 $k_1\ll  k_2\sim k_3$ of  the three-point function is particularly interesting from the
 observationally point of view because it is associated to the simplest model of NG, the so-called local one in
which the total initial adiabatic curvature is a local function of its gaussian counterpart $\zeta_{\rm g}$, {\it e.g.}
$\zeta=\zeta_{\rm g}+3 f_{\rm NL}/5(\zeta_{\rm g}^2-\langle \zeta_{\rm g}^2\rangle)+\cdots$, where  $f_{\rm NL}$ the nonlinear coefficient parametrizing the amplitude of NG
 (for a review see  \cite{revNG}). The local model leads to pronounced effects of NG on the clustering of dark matter halos and to  strongly scale-dependent bias \cite{dalal}. 
 
In the squeezed limit 
the three-point correlator (\ref{3pkc}) for $X\sim Y\gg1$ turns out to be
\be
&& \langle\sigma^I_{\vk_1}\sigma^J_{\vk_2}\sigma^K_{\vk_3}\rangle^\prime= 
c_{IJK}\, \gamma_s\, \frac{1}{k_1^{3-\Delta_I-\Delta_J}}
\frac{1}{k_2^{3-\Delta_K}}+\,{\rm cyclic}, \label{kk1}
\ee
where 
\be
\gamma_s=(2\pi)^3 \,\, 
\frac{ 
\Gamma
  \left(\frac{3}{2}\!-\!
\frac{\Delta_K}{2}\right)
}{2^{3(\Delta-1)}
\Gamma
  \left(\frac{\Delta_K}{2}\right)}     
\frac{\Gamma
  \left(\frac{3}{2}\!-\!\frac{\Delta_K}{2}\!-\!\frac{3 \Delta}{2} \right)}{ \Gamma
   \left(\frac{\Delta_I+\Delta_J}{2}\right)}.
\ee
For the case of scalars of equal dimensions $\Delta_I=\Delta_J=\Delta_K=w$, Eq. (\ref{3pkc}) is written as 
\be
\langle\sigma^I_{\vk_1}\sigma^J_{\vk_2}\sigma^K_{\vk_3}\rangle^\prime 
&=&c_{IJK}2^{7-3\Delta}\pi^{\frac{5}{2}}
\frac{\Gamma(3-\frac{3w}{2})\Gamma(\frac{3-w}{2})}{\Gamma(\frac{w}{2})^2}\nonumber \\
&\times&
k_1^{3w-6}\int_0^1 {\rm d}u\frac{(1-u)^{\frac{w}{2}-1}u^{\frac{w}{2}-1}}{
\left[(1\!-\!u) X\!+\!u Y\right]^{\frac{3-w}{2}}} \, \,{}_2F_1\!\!\left(\frac{3w}{2}\!-\!\frac{3}{2},\frac{3-w}{2},
\frac{3}{2},{\cal{Z}}\right)+\,{\rm cyclic},
\label{3pkcm}
\ee
which coincides with the corresponding expression in Ref. \cite{anto} after the Euler's transformation (\ref{euler}) in order 
the hypergeometric function to converge in $|{\cal{Z}}|\leq 1$.

 Applying the squeezed limit  
 to the expression   (\ref{3pkcm}) we find that  the generic NG three-point function has the form 
\be
 \langle\sigma^I_{\vk_1}\sigma^J_{\vk_2}\sigma^K_{\vk_3}\rangle^\prime\sim 
\, \gamma_s\, \frac{c_{IJK}}{k_1^{3-2w} k_2^{3-w}}
 +{\rm cyclic}\,\,\,\,\,\,\,\,\,\,\,(k_1\ll k_2\sim k_3). \label{kk2}
\ee
This result  is  dictated by simple scale invariance and not by full conformal symmetry and fixes the
shape of the three-point configuration in the squeezed limit up to a model-dependent
coefficient $c_{IJK}$.
We note that the result (\ref{kk2}) does not coincide with the squeezed limit found in Ref. \cite{anto} for the generic three-point function in the squeezed limit.
The reason is that, the authors of Ref. \cite{anto} took the squeezed limit of (\ref{3pkcm}) before  
Euler transforming the hypergeometric function in (\ref{3pk})  which is  not defined for 
scalar masses within the unitarity bounds $0\leq m_{I,J,K}<3/2 H$. Here, after Euler transforming,  the integral 
in (\ref{3pkcm}) is well defined
for masses in the 
unitarity region and the expected behaviour (\ref{kk2}) is recovered. 


If one is interested just in the squeezed limit of the three-point correlator, this may be found quite easily in an another way. Let us consider again, for 
simplicity, the case $\Delta_I=\Delta_J=\Delta_K=w$ so that the three-point function  (\ref{3p}) is written as
\be
\langle\sigma^I(\vx_1)\sigma^J(\vx_2)\sigma^K(\vx_3)\rangle=
\frac{c_{IJK}}{|\vx_1-\vx_2|^{w}|\vx_2-\vx_3|^{w}|\vx_3-\vx_1|^{w}}.\label{3ps1}
\ee
In the limit $x_{23}\simeq 0 \, , ~x_{13}=x_{12}\gg x_{23}$ where $x_{ij}=|\vx_i-\vx_j|$,  
 (\ref{3ps1}) may be expressed as 
 \be
\langle\sigma^I(\vx_1)\sigma^J(\vx_2)\sigma^K(\vx_3)\rangle=
\frac{c_{IJK}}{x_{23}^{w}x_{13}^{2w}}.\label{3ps2}
\ee
By using   
\be
\frac{1}{|\vx|^{w}}=\frac{\Gamma(\frac{3-w}{2})}{2^{w}\pi^{3/2}
\Gamma(\frac{w}{2})}\int {\rm d}^3 k\,|\vk|^{w-3}e^{-i\vk\cdot \vx}\, ,\label{xw}
\ee
for each factor in the denominator of (\ref{3ps2})
and Fourier transforming it, we get 
\be
\langle\sigma^I_{\vk_1}\sigma^J_{\vk_2}\sigma^K_{\vk_3}\rangle\propto
\int {\rm d}^3\vx_1{\rm d}^3\vx_2 {\rm d}^3\vx_3e^{i \vk_1\cdot \vx_1+i\vk_2\cdot \vx_2+i\vk_3\cdot \vx_4}
\int {\rm d}^3\vq_1 {\rm d}^3\vq_2 \frac{e^{-i\vq_1\cdot (\vx_2-\vx_3) -i\vq_2\cdot (\vx_1-\vx_3)}}{|\vq_1|^{3-w}|\vq_2|^{3-2w}}.
\ee
The integration of space  specify the external momenta $\vk_i$ are given by
\be
\vk_1=\vq_2\, , ~~ \vk_2=\vq_1\, , ~~ \vk_1=\vq_1+\vq_2\, . 
\ee
Finally, integration over the internal momenta $\vq_i$, specifies the three-point function in the squeezed limit to be
\be
   \langle\sigma^I_{\vk_1}\sigma^J_{\vk_2}\sigma^K_{\vk_3}\rangle^\prime
\sim  
\frac{1}{k_1^{3-2w} k_2^{3-w}}. \label{kks}
\ee
This is what we found above in Eq. (\ref{kk2}) by using the exact expression of the three-point correlator. Now, since $|\vq_1|\gg |\vq_2|$ as $\vq_1, ~\vq_2$ are conjugate momenta of $x_{23}$
and $x_{13}=x_{12}$ (with $x_{23}\to 0 \, , ~x_{13}=x_{12}\gg x_{23}$), we get that (\ref{kks}) is valid in the $k_1\ll k_2\sim k_2$. 
In other words, the local shape of the three-point function corresponds to two point close and one remote point in three-dimensional 
space as shown in Fig. 1.  
\begin{figure}[h!]
\begin{center}
 \includegraphics[scale=0.5]{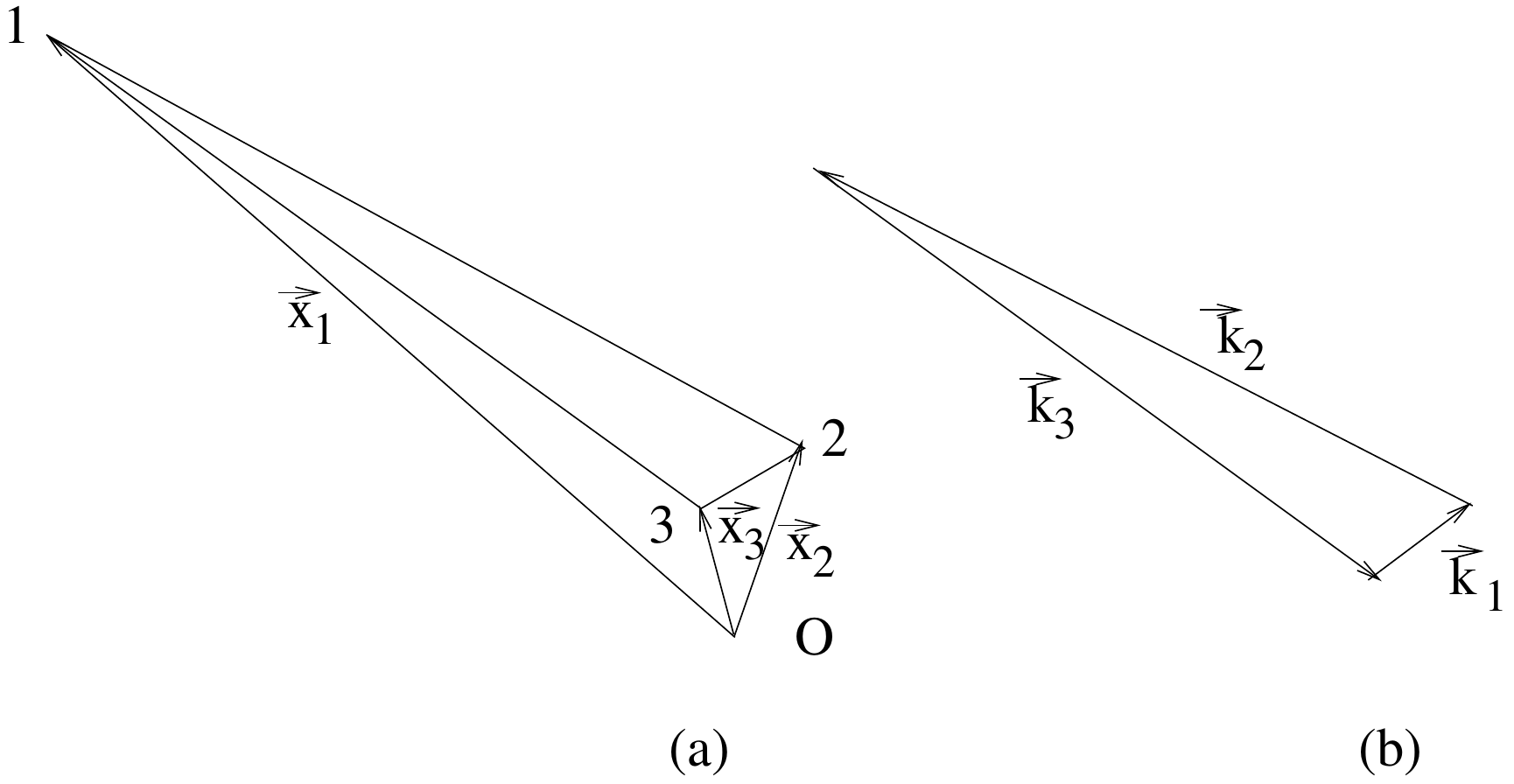}
\caption{(a) Squeezed three-point  configuration with two points  $\vx_2$ and $\vx_3$  very close (O being the origin) and the third one $\vx_1$ 
far away from the 
rest, {\it i.e.}  $x_{23}\simeq 0 \, , ~x_{13}=x_{12}\gg x_{23}$. (b) Local shape
in $k$-space  with $k_1\ll k_2\sim k_2$.}
\end{center}
\end{figure} 
In the equilateral case $k_1=k_2=k_3=k$, we have  $X=Y=1$ and ${\cal{Z}}=1-u(1-u)$ and thus
\be
 \langle\sigma^I_{\vk_1}\sigma^J_{\vk_2}\sigma^K_{\vk_3}\rangle^\prime\sim 
c_{IJK}\, \gamma_e\, k^{3\Delta-6},
\ee
where 
\be
&&\gamma_e=2^{10-3\Delta}\pi^{\frac{11}{2}}\frac{\Gamma(3-\frac{3\Delta}{2})\Gamma(\frac{3-\Delta_K}{2})}{\Gamma(\frac{\Delta_I}{2})
\Gamma(\frac{\Delta_J}{2})}\times \nonumber \\
&&\hspace{1.2cm} \int_0^1 {\rm d}u\, (1-u)^{\frac{1}{2}-\frac{\Delta_I}{2}}u^{\frac{1}{2}-\frac{\Delta_J}{2}} \, 
\, {}_2F_1\!\!\left(3\!-\!\frac{3\Delta}{2},\frac{\Delta_K}{2},\frac{3}{2},1\!-\!u(1\!-\!u)\right).
\ee
Finally, let us comment on the massless limit 
\be
w_I=\frac{3}{2}\left(1-\sqrt{1-\frac{4m_I^2}{9H^2}}\right)\ll 1. 
\ee
The  two- and three-point functions are to be obtained  in the 
 $\Delta_I\sim 0$ limit. This limit can be smoothly found by employing Eq. (\ref{xw}), which  
expanded around $w_{I}=0$, gives 
\be
&&\hspace{-.5cm}\lim_{w_{I}\to 0}\frac{1}{|\vx|^{w_{I}}}= \lim_{w_{I}\to 0} \frac{\Gamma(\frac{3-w_{I}}{2})}{2^{w_{I}}\pi^{3/2}\Gamma(\frac{w_{I}}{2})}\int {\rm d}^3 k\,
|\vk|^{w_{I}-3}e^{-i\vk\cdot \vx}
 \nonumber \\
&&\hspace{1.4cm} =\!\frac{w_{I}}{4\pi}\!\left\{1\!+\!\frac{w_{I}}{2}\left[\gamma\!-\!2\ln 2\!-\!\psi\left(\frac{3}{2}\right)\right]
\right\}\!
\Big{(}\!\ln|\vx|\!-\!\frac{1}{2}\gamma\Big{)}
\!+\!\frac{w_{I}^2}{8\pi}\!\left(\ln^2|\vx|\!-\!\gamma\ln |\vx|\!+\!\frac{1}{24}(6\gamma\!+\!\pi^2)\!\right),\nonumber\\
&&
\ee
where we have used 
\be
\int {\rm d}^3 k\, e^{i \vk\cdot\vx} \frac{4\pi}{|\vx|^{2n+3}}=\frac{(-1)^n\pi^{3/2}}{n!2^{2n}\Gamma(n+\frac{3}{2})}\left(
\phantom{\frac{2^2}{2^2\frac{1}{2}}}\!\!\!\!\!\!\!\!\!\!
|\vx|^{2n}
\ln|\vx|+\psi(n+1)\right).
\ee
As a result, the most singular behaviour of the two- and three-point functions in the 
$w_{I}\approx m_I^2/3H^2\ll 1$ limit is
\be
&&\langle \sigma^I(\vx_1)\sigma^I(\vx_2)\rangle \sim \ln |\vx_1-\vx_2|\, ,\label{2pss} \\
&&\langle \sigma^I(\vx_1)\sigma^I(\vx_2)\sigma^I(\vx_3)\rangle\sim
 \ln |\vx_1-\vx_2|\,  \ln |\vx_1-\vx_3|\,  \ln |\vx_3-\vx_2|. \label{3pss}
\ee

\subsection{Conformal Invariance}
\noindent
Let us now enhance the symmetry by demanding also invariance under special conformal transformations generated by $K_i$. In this 
case, the symmetry turns out to be the full conformal invariance generated by rotations, translations, dilitations and special conformal
transformations. Correlators are more restricted now as special conformal transformations gives additional conditions. Since, special 
conformal transformations are reduced to inversion, it is enough to consider just transformations under space inversion (\ref{inv}).
Let us recall that under conformal transformations, the two-point function of fields $\sigma^I$ and $\sigma^J$ of 
conformal dimensions $\Delta_I$ and $\Delta_J$ respectively, transforms as 
\be
\langle\sigma^I(\vec{{x}}_1)\sigma^J(\vx_2)\rangle\to \Big{|}\frac{\partial x_i'}{\partial x_j}\Big{|}_{x=x_1}^{\Delta_I/3} 
\Big{|}\frac{\partial x'_i}{\partial x_j}\Big{|}_{x=x_2}^{\Delta_J/3} \langle\sigma^I(\vec{{x}}'_1)\sigma^J(\vx_2')\rangle
\ee
where $|\partial x_i' /\partial x_j|$ is the Jacobian of the transformation. For the  space inversion (\ref{inv}),
the two-point function  (\ref{2p}), the form of which was forced by scale invariance,  transforms
as
\be
\langle\sigma^I(\vec{{x}}_1)\sigma^J(\vx_2)\rangle\to \frac{(r_1 r_2)^{\Delta_I+\Delta_J}}{r_1^{2\Delta_I}r_2^{2\Delta_J}}
\langle\sigma^I(\vec{{x}}_1)\sigma^J(\vx_2)\rangle,
\ee
 where for $\vx'=\vx/|\vx|^2$ we have used that   
\be
\Big{|}\frac{\partial x_i'}{\partial x_j}\Big{|}=\frac{1}{|\vx|^6}\, , ~~~|\vx_1-\vx_2|\to \frac{|\vx_1-\vx_2|}{r_1^2 r_2^2},
\label{spinv}
\ee
and the notation $|\vx_1|=r_1,|\vx_2|=r_2$. Thus, space inversion leaves the two point function invariant if 
\be
\Delta_I=\Delta_J.
\ee
Similarly,  the three-point function transforms as 
\be
\langle\sigma^I(\vx_1)\sigma^J(\vx_2)\sigma^K(\vx_3)\rangle\to
\Big{|}\frac{\partial x_i'}{\partial x_j}\Big{|}_{x=x_1}^{\Delta_I/3} 
\Big{|}\frac{\partial x_i'}{\partial x_j}\Big{|}_{x=x_1}^{\Delta_J/3} 
\Big{|}\frac{\partial x_i'}{\partial x_j}\Big{|}_{x=x_1}^{\Delta_K/3} 
 \langle\sigma^I(\vx_1')\sigma^J(\vx_2')\sigma^K(\vx_3')\rangle
\ee
and using (\ref{spinv}), we get that (\ref{3p}) is invariant if 
\be
w_K=\Delta_I+\Delta_J-\Delta_K\, ,~~w_I=\Delta_J+\Delta_K-\Delta_I\, ,~~w_J=\Delta_I+\Delta_K-\Delta_J.
\ee 
As a result,  two- and three-point function are conformal invariant if they have the form
\be
\hspace{-1cm}&&\langle\sigma^I(\vec{{x}}_1)\sigma^J(\vx_2)\rangle=\left\{\begin{array}{cl}
                                                           \frac{c_{IJ}}{|\vx_1-\vx_2|^{2\Delta_I}}
&\Delta_I=\Delta_J,\label{2pc}\\
0 & \Delta_I\neq \Delta_J,
                                                          \end{array}
\right.
 \\
\hspace{-1cm}&&\langle\sigma^I(\vx_1)\sigma^J(\vx_2)\sigma^K(\vx_3)\rangle=
\frac{c_{IJK}}{|\vx_1-\vx_2|^{\Delta_I\!+\!\Delta_J\!-\!\Delta_K}|\vx_2-\vx_3|^{\Delta_J\!+\!\Delta_K\!-\!\Delta_I}
|\vx_3-\vx_1|^{\Delta_I\!+\!\Delta_K\!-\!\Delta_J}}, \label{3pc}
\ee
where again $\sigma^{I,J,K}$ are operators of dimensions $\Delta_{I,J,K}$. In other words, enhancing the symmetry including the special conformal symmetry has two consequences. First, the two-point functions are zero for 
operators with different dimensions and, second, the three-point functions are completely specified by special conformal transformations, {\it i.e.} by the full conformal symmetry.

We deduce  that   in multi field models,  conformal symmetry imposes that the scalar fields are uncorrelated at the level of two-point correlators if their masses are different. This seems to have passed unnoticed in the recent literature on the subject and   usually the fact that the $\sigma^I$ fields are not correlated is taken as an assumption. We see that in fact it is a consequence of the conformal symmetry:  if  the conformal weights of two light fields are different, then their cross-correlation vanishes:

\be
\fbox{$\displaystyle
{\rm If } \,\,\, \Delta_I\neq \Delta_J \Rightarrow \langle\sigma^I(\vx_1)\sigma^J(\vx_2)\rangle= 0$}.
\ee
Although this is a classical result, quantum corrections may induce anomalous dimensions to the fields. 
However, as long as two fields have 
different dimensions at some order in perturbation theory (or even non-perturbatively), 
their two-point function vanishes by conformal invariance at that order. 
On the other hand, it may happen that two fields  have 
the same dimension at the classical level, for example due to the same tree-level mass. 
However, if there is no  symmetry to protect this tree-level 
relation, interactions will spoil it by introducing different  dimensions to the fields. In this case, although
classically they have a non-zero two-point correlator, the latter will vanish at the quantum level 
(as long as there is no conformal anomaly).  

\section{The operator product expansion and the NG  correlators}
\noindent
After this excursion on the  the symmetries present in a de Sitter geometry and the constraints they impose on the two-and three-point 
correlators   of light fields and their shapes, let us proceed with the more original part of the work and consider the informations we can gather using the Operator Product Expansion (OPE) for fields considered at  coincidence points when the system enjoys the symmetries of the de Sitter geometry. As we shall see, 
the OPE is particularly useful and powerful
to characterize in their full generality the squeezed limit of the three-point correlator and the collapsed limit of the
fout-point correlator which are    interesting limits from the observationally point of view.
 
The OPE  has been established in perturbative quantum field theories. 
It is by now a standard tool in the analysis of 
gauge theories such as QCD and  Wilson's OPE \cite{Wilson} is the basis of virtually all calculations of nonperturbative effects in
analytical QCD.  
It is believed that all quantum field theories with well-behaved ultraviolet behavior  have
an operator product expansion (OPE) \cite{Wilson,wilson,zim}. 
This has  been proven for conformally invariant quantum
field theories  \cite{ope1,ope2}. In particular, OPE  in two-dimensional
conformal field theories has  played a major role in the development of string theory 
and critical phenomena \cite{ope3}. In addition,  in search for a more solid foundation of 
OPE expansion, formal mathematical proofs of its existence and validity have also been
given within various axiomatic settings \cite{ope4} for quantum field theory on Minkowski
spacetime. There are also formulations of the operator product expansion of local operators in curved spacetimes \cite{ope5}.

Let us consider two generic operators $\sigma^I(\vx)$ and $\sigma^J(\vy)$ at the points $\vx$ and $\vy$ on a 
$\eta={\rm const.}$ hypersurface of de Sitter spacetime. 
Then, we expect that the product of  local operators are distances small compared to the 
characteristic length of the system should look like a local operator. As a result, we expect that
 the product of $\sigma^I(\vx)\sigma^J(\vy)$ of the two operators $\sigma^I(\vx)$ and $\sigma^J(\vy)$, located at nearby points $\vx$ and $\vy$, will have a 
short-distance expansion of the form  \cite{Wilson}
\be
\sigma^I(\vx) \sigma^J(\vy)\stackrel{\vx\to\vy}{\sim}\sum_nC_{n}(\vx-\vy){\cal{O}}_n(\vy),  \label{ab1}
\ee
where $C_n(\vx-\vy)$ are c-number functions (in fact distributions) and ${\cal{O}}_n$ local operators.                                                                                                  
Moreover, for $H\eta \ll 1$ we expect the OPE (\ref{ab1}) to respect the symmetries of the de Sitter
spacetime realized non-linearly on the $\eta={\rm const.}$ hypersurface. In other words, we expect (\ref{ab1}) to 
enjoy conformal three-dimensional symmetry. Note that the OPE above  
 can also be written as  
\be
\sigma^I(\vx)\sigma^J(\vec{0}){\displaystyle\stackrel{\vx\to\vec{0}}{\sim}}
\sum_{n,s}C_{ns}(\vx)x^{i_1}x^{i_2}\cdots x^{i_s}{\cal{O}}^{(ns)}_{i_1i_2\cdots i_s}(\vec{0}) \label{OPE1}
\ee
since due to translational invariance we have taken $\vy=\vec{0}$ and 
\be
{\cal{O}}_n(\vx)=e^{iP_ix^i}{\cal{O}}_n(\vec{0})e^{-iP_ix^i}=\sum_s
\frac{1}{s!}x^{i_1}x^{i_2}\cdots x^{i_s}{\cal{O}}^{(ns)}_{i_1i_2\cdots i_s}(\vec{0}),
\ee
where 
\be
{\cal{O}}^{(ns)}_{i_1i_2\cdots i_s}(\vec{0})=(-i)^n[P_{i_1},[P_{i_2},\cdots [P_{i_s},{\cal{O}}^{(n)}(\vec{0})],\cdots]].
\ee
We assume now  that the local operators ${\cal{O}}^{(ns)}_{i_1\ldots i_s}(\vx)$ transforms under rotations,
 translations and dilations as \cite{fgg0}
\be
&&i [L_{ij},{\cal{O}}^{(ns)}_{i_1\ldots i_s}]=\left(x_i\partial_j-x_j\partial_i+i\Sigma_{ij}^{(s)}\right){\cal{O}}^{(ns)}_{i_1\ldots i_s}, \\
&&i[D,{\cal{O}}^{(ns)}_{i_1\ldots i_s}]=\left(x^i\partial_i+\Delta_s\right){\cal{O}}^{(ns)}_{i_1\ldots i_s},\label{ddee} \\
&&i[K_i,{\cal{O}}^{(ns)}_{i_1\ldots i_s}]=\left(2\Delta x_i+2x_i x^j \partial_j-x^2\partial_i+2i x^j\Sigma_{ji}^{(s)}\right)
{\cal{O}}^{(ns)}_{i_1\ldots i_s},\label{kee} \\
&&i[P_i,{\cal{O}}^{(ns)}_{i_1\ldots i_s}]=\partial_i {\cal{O}}^{(ns)}_{i_1\ldots i_s}, 
\ee
where, as before $\Sigma_{ij}^{(s)}, \Delta,K_i$ are representations of the stability group at $\vx=\vec{0}$ and 
the index $n$ just labels the representations of the latter. The operators ${\cal{O}}^{(nn)}_{i_1\ldots i_s}(\vx)$
are the lowest dimensional operator (of dimension $w_n$), they  commute with $K_i$ and are  the primary fields in the theory. 
The action of $\Delta$ on a representation 
${\cal{O}}^{(ns)}_{i_1\ldots i_s}(\vec{0})$ is 
\be
\label{dde}
[{\cal{O}}^{(ns)}_{i_1\ldots i_s}(\vec{0}),\Delta]=i(w_n+m-n){\cal{O}}^{(ns)}_{i_1\ldots i_s}(\vec{0}).
\ee
For operators $A$ and $B$ of dimensions $w_I$ and $w_J$ respectively,  scale invariance provides  the condition
\be
[\sigma^I(\vx)\sigma^J(\vec{0}),D]=(x^i\partial_i+w_I+w_J)\sigma^I(\vx)\sigma^J(\vec{0}).
\ee
By employing Eqs. (\ref{ddee}) and (\ref{dde}), we obtain the following equation 
for $C_{ns}(\vec{x})$
\be
x^i\partial_i C_{ns}+(w_n-n+w_I+w_J)C_{ns}=0.
\ee
Thus, scale invariance specifies $C_{ns}\sim x^{w_n-n-w_I-w_J}$, so that Eq. (\ref{OPE1}) turns out to be
\be
\sigma^I(\vx)\sigma^J(\vec{0})\stackrel{\vx\to\vec{0}}{\sim}\sum_{n}\left(\frac{1}{|\vx|}\right)^{w_I+w_J-w_n+n}\sum_s C_{ns}x^{i_1}x^{i_2}
\cdots x^{i_s}{\cal{O}}^{(ns)}_{i_1i_2\cdots i_s}(\vec{0}). \label{OPE2}
\ee
The contribution of the scalar $s=0$ sector is, for example,  
\be
\sigma^I(\vx)\sigma^J(\vec{0})\stackrel{\vx\to\vec{0}}{\sim}\left(\frac{1}{|\vx|}\right)^{w_I\!+\!w_J}\left\{C_0+
\phantom{(\frac{1}{|\vx|}}
\hspace{-.6cm}|\vx|^{-w_{\cal{O}}}\left(\phantom{(\frac{1}{|\vx|}}
\hspace{-.6cm}C_1{\cal{O}}(\vec{0})+C_1 x^i\partial_i {\cal{O}}(\vec{0})+\ldots\right)\right\}, \label{abvx}
\ee
where the dots stand for the contributions of higher spin descendants.
By demanding simply scale invariance, we have arrived in the short-distance expansion (\ref{OPE2}). We may continue and investigate the
constraints which full conformal symmetry  further imposes. To do that, we have just to impose invariance 
under special conformal 
transformations. Although it is a straightforward process, it is quit involved and we point out here only the main steps
repeating basically the corresponding steps taken by Ferrara, Gatto and Grillo in  Refs.\cite{fgg1,fgg2} for the $SO(2,4)$ case. Commuting 
(\ref{OPE2}) with the generator of special conformal transformations, we get
\be
[\sigma^I(\vx)\sigma^J(\vec{0}),K_i]=-i\sum_{n}\left(\frac{1}{|\vx|}\right)^{w_I+w_J-w_n+n}\sum_s C_{ns}x^{i_1}x^{i_2}
\cdots x^{i_s}\, [{\cal{O}}^{(ns)}_{i_1i_2\cdots i_s}(\vec{0}),K_i].
\ee
Then, by using Eqs. (\ref{sigma}) and (\ref{kee}), we get 
\be
&&\hspace{-1.5cm}[\sigma^I(\vx)\sigma^J(\vec{0}),K_i]=-i\sum_{n=0}\!\!\left(\frac{1}{|\vx|}\right)^{w_I\!+\!w_J\!-\!w_n\!+\!n}\!\!\sum_{s=n} C_{ns}
(w_I\!+\!2s\!-\!w_J\!+\!n\!-\!w_o)x_ix^{i_1}x^{i_2}
\cdots x^{i_s}{\cal{O}}^{(ns)}_{i_1i_2\cdots i_s}(\vec{0})\nonumber\\
&&\hspace{1cm}=-i\sum_{n=0}\left(\frac{1}{|\vx|}\right)^{w_I\!+\!w_J\!-\!w_n\!+\!n}\sum_{s=0} C_{ns}(s\!+\!1\!-\!n)
(w_n\!+\!s\!+\!1)
x_ix^{i_1}x^{i_2}
\cdots x^{i_s}{\cal{O}}^{(ns)}_{i_1i_2\cdots i_s}(\vec{0}).
\ee
From the above relation, we obtain the recurrence equation
\be
2k(w_n+n+k)C_{n,n+k}-(w_I-w_J+w_n+n-2k-2)C_{n,n+k}=0,
\ee
which is solved by 
\be
C_{n,n+k}=\frac{\Gamma\left(\frac{1}{2}(w_I-w_J+n)+k\right)\Gamma(w_n+n)}{k!
\Gamma\left(\frac{1}{2}(w_I-w_J+n)\right)\Gamma(w_n+n+k)}C_{n,n}.
\ee
By recalling that 
\be
&&\sum_{m=n}^\infty C_{nm}z^{m-n} =\sum_{k=0}^\infty C_{n,n+k} z^k= \sum_{k=0}^\infty \frac{\Gamma\left(\frac{1}{2}(w_I-w_J+n)+k\right)\Gamma(w_o+n)}{k!
\Gamma\left(\frac{1}{2}(w_I-w_J+n)\right)\Gamma(w_o+n+k)}C_{n,n}z^n \nonumber \\
&&\hspace{2.5cm} =
\Phi\left(\frac{1}{2}(w_I-w_J+n);w_n+n;z\right),
\ee
we get finally \cite{fgg1,fgg2}
\be
\sigma^I(\vx)\sigma^J(\vec{0})\stackrel{\vx\to \vec{0}}{\sim}\sum_{n}C_n\frac{1}{x^{\ell_1+\ell_2}}x^{i_1}x^{i_2}
\cdots x^{i_n}\, \Phi\!\left(\!\frac{1}{2}(\ell_1\!-\!\ell_2);w_n\!+\!n;x^i\partial_i\right)
{\cal{O}}_{i_1i_2\cdots i_n}(\vec{0}). \label{OPE3}
\ee
We have indicated  here 
\be
\ell_1=w_I+n\, , ~~~\ell_2=w_J-w_n, 
\ee
and $\Phi\left(\!\frac{1}{2}(w_I\!-\!w_J\!+\!w_n\!+\!n);w_n\!+\!n;x^i\partial_i\right)$ is the confluent hypergeometric function 
defined  as
\be
\Phi(a;b;x^i\partial_i)=\sum_n \frac{(a)_n}{(b)_n} (x^i\partial_i)^n.
\ee
Moreover, $(a)_n,(b)_n$ are the Pochhammer symbols for $a=\frac{1}{2}(w_I-w_J+w_n+n)$ and 
$b=w_n+n$. 
By using the integral representation of the confluent hypergeometric function 
\be
\Phi(a;b;z)=\frac{\Gamma(b)}{\Gamma(b-a)\Gamma(a)}\int_0^1 {\rm d}u \,u^{a-1}(1-u)^{b-a-1}e^{u z} \label{confint}
\ee
and 
\be 
e^{u x^i\partial_i}f(\vec{0})=f(ux^i),
\ee 
we may express the OPE (\ref{OPE3}) as 
\be
\sigma^I(\vx)\sigma^J(\vec{0})\stackrel{\vx\to\vec{0}}{\sim}\sum_{n}C_n\frac{1}{x^{\ell_1+\ell_2}}x^{i_1}x^{i_2}
\cdots x^{i_n}\, \int \,{\rm d}^3k \,  \Phi\!\left(\!\frac{1}{2}(\ell_1\!-\!\ell_2);w_n\!+\!n;i\vk\cdot\vx\right)
\widetilde{{\cal{O}}}_{i_1i_2\cdots i_n}(\vk), \label{OPE33}
\ee
where $\widetilde{{\cal{O}}}_{i_1i_2\cdots i_n}(\vk)$ is the Fourier transform of ${{\cal{O}}}_{i_1i_2\cdots i_n}(\vx)$.
It should be stressed that the OPE also determines the structure of the $n$-point functions. 
For instance, the two point-function of two scalar operators $\sigma^I$ and $\sigma^J$ of dimensions $w_I$ and $w_J$ respectively
 is given by Eq. (\ref{OPE3}) as 
\be
&&\langle \sigma^I(\vx)\sigma^J(\vec{0})\rangle\sim\sum_{n} C_n \frac{1}{x^{\ell_1+\ell_2}}x^{i_1}x^{i_2}
\cdots x^{i_n}\,\int \,{\rm d}^3k \,  \Phi\!\left(\!\frac{1}{2}(\ell_1\!-\!\ell_2);w_n\!+\!n;i\vk\cdot\vx\right)
\langle\widetilde{{\cal{O}}}_{i_1i_2\cdots i_n}(\vk)\rangle  \nonumber \\
&&
\phantom{\langle \sigma^I(\vx)\sigma^J(\vec{0})\rangle}=\frac{1}{x^{w_I+w_J}}\,\int {\rm d}^3k \,  
\Phi\!\left(\!\frac{1}{2}(w_J-w_I);0;i\vk\cdot\vx\right)
\langle\widetilde{{\cal{O}}}(\vk)\rangle,\label{OPE44}
\ee
since from rotational invariance of the vacuum, only $SO(3)$ singlets will contribute to the 
right-hand side of Eq. (\ref{OPE44}), 
that is only operators for which  $n=0$ and $w_{\cal O}=0$ operators.  
Then, it is clear that only for $w_I\neq w_J$, the integral in the right-hand side of Eq. (\ref{OPE44}) is a function of $\vx$. However, 
by scale invariance, the integral should be $\vx$-independent  and,  in fact, it can only be a numerical constant. This is however possible
only for $w_I=w_J$ for which   $\Phi\!\left(0;0;i\vk\cdot\vx\right)=1$. 
For $w_I\neq w_J$, $\langle\widetilde{{\cal{O}}}\rangle=0$ for a dimensionful operator since 
otherwise conformality will be lost. As  result, with $C_{\cal O}$ the coefficient of the identity operator in the operator product 
expansion, we get 
\be
\langle \sigma^I(\vx)\sigma^J(\vec{0})\rangle= \left\{\begin{array}{cl}
                                                          \frac{C_{\cal O}}{x^{w_I+w_J}}
&w_I=w_J,\label{2pc1}\\
0 & w_I\neq w_J.
                                                          \end{array}
\right.
\ee
{\it i.e.} we recover  Eq. (\ref{2pc}) as expected. 

\subsection{The three-point function from the OPE and its  squeezed limit}
\noindent
Similar considerations also may apply to three-point, or generally to $n$-point correlators. 
By employing the integral representation (\ref{confint}) of the confluent hypergeometric function in the OPE (\ref{OPE3}), we get
\be
&&\sigma^I(\vx)\sigma^J(\vec{0})\stackrel{\vx\to\vec{0}}\sim\sum_{n}C_n\frac{1}{x^{\ell_1+\ell_2}}x^{i_1}x^{i_2}
\cdots x^{i_n}\int_0^1 {\rm d}u\, u^{a-1}(1-u)^{b-a-1} {\cal{O}}_{i_1\cdots i_n}(u\vx).
\ee
This form of the OPE is quite appropriate to calculate the three-point correlator $\langle \sigma^I(\vx)\sigma^J(\vec{0})\sigma^K(\vy)\rangle$ 
of three  scalar operators $A$, $B$ and $C$, with dimensions $w_I$, $w_J$ and $w_C$ respectively.    
We find 
\be
\langle \sigma^I(\vx)\sigma^J(\vec{0})\sigma^K(\vy)\rangle\stackrel{\vx\to\vec{0}}{\sim}\sum_{n}C_n\frac{1}{x^{\ell_1+\ell_2}}x^{i_1}x^{i_2}
\cdots x^{i_n}\int_0^1 {\rm d}u \, u^{a-1}(1-u)^{b-a-1}\,  \langle{\cal{O}}_{i_1\cdots i_n}(u\vx) \, \sigma^K(\vy)\rangle \label{coo}.
\ee
We now use the orthogonality of the two-point correlator.  This means, in particular, 
that only scalar operators ${\cal{O}}$ ($n=0$) will contribute to 
the right-hand side of Eq. (\ref{coo})  
\be
\langle \sigma^I(\vx)\sigma^J(\vec{0})\sigma^K(\vy)\rangle\stackrel{\vx\to\vec{0}}{\sim} C_{\cal O}\frac{1}{x^{w_I+w_J-
w_{\cal O}}}\ \int_0^1 {\rm d}u \, u^{a-1}(1-u)^{w_{\cal O}-a-1}\langle{\cal{O}}(u\vx)\sigma^K(\vy)\rangle.
\label{abc}
\ee
The two-point function  $\langle{\cal{O}}(u\vx)\sigma^K(\vy)\rangle$ is given by
\be
\langle{\cal{O}}(u\vx)\sigma^K(\vy)\rangle\sim \left\{\begin{array}{cc}
                                   \frac{1}{|\vy-u\vx|^{2w_C}}&w_{\cal O}=w_C,\\
0&w_{\cal O}\neq w_C.
                                  \end{array}
\right.
\ee  
and Eq. (\ref{abc}) turns out to be 
\be
\langle \sigma^I(\vx)\sigma^J(\vec{0})\sigma^K(\vy)\rangle\stackrel{\vx\to\vec{0}}{\sim}\frac{1}{x^{w_I+w_J-w_C}} \int_0^1 {\rm d}u\, \frac{u^{a-1}(1-u)^{w_{\cal O}-a-1}}{(y^2
-2u \,\vx\cdot \vy)^{w_C}}. 
\label{abc1}
\ee
Using  Feynman parameters
\be
 \int_0^1 {\rm d}u\, \frac{u^{\alpha_1-1}(1-u)^{\alpha_2-1}}{(u D_1+(1-u)D_2)^{\alpha_1+\alpha_2}}=
\frac{\Gamma(\alpha_1)\Gamma(\alpha_2)}{
\Gamma(\alpha_1+\alpha_2)}
\frac{1}{D_1^{\alpha_1}D_2^{\alpha_2}}, 
\ee
we have
\be
&&\hspace{-1cm}\int_0^1 {\rm d}u \frac{u^{a-1}(1-u)^{w_{\cal O}-a-1}}{(y^2
-2u\, \vx\cdot \vy)^{w_C}} =\int_0^1 {\rm d}u \frac{u^{a-1}(1-u)^{w_{\cal O}-a-1}}{
\big{\{}u\,(y^2-2u\, \vx\cdot \vy)+(1-u)y^2\big{\}}^{w_C}}\nonumber \\
&&\hspace{3.3cm} =\frac{\Gamma(w_C-a)\Gamma(a)}{\Gamma(w_C)}\frac{1}{(y^2-2u\, \vx\cdot \vy)^{a}}\frac{1}{y^{2(w_C-a)}},
\ee
and thus, the three-point function as specified by the OPE (\ref{abc1}) is found to be 
\be
\langle \sigma^I(\vx)\sigma^J(\vec{0})\sigma^K(\vy)\rangle\stackrel{\vx\to\vec{0}}{\sim} \frac{1}{|\vx|^{w_I+w_J-w_C}} \frac{1}{|\vy-\vx|^{w_I-w_J+w_C}}\frac{1}{|\vy|^{w_J+w_C-w_I}}.
\label{abcf}
\ee
The OPE's are particularly appropriate for calculating the squeezed limit of correlators. Let us consider 
for 
example a single scalar whose two-point correlator is 
\be
\langle\sigma(\vx)\sigma(\vy)\rangle\sim\frac{1}{|\vx-\vy|^{2w}}. \label{ph}
\ee
Let us  consider
now the OPE 
\be
\sigma(\vx)\sigma(\vec{0})\stackrel{\vx\to\vec{0}}{\sim}\sum_{n}C_n\frac{1}{x^{\ell_1+\ell_2}}x^{i_1}x^{i_2}
\cdots x^{i_n}\, \int {\rm d}^3k \,  \Phi\!\left(\!\frac{1}{2}(\ell_1\!-\!\ell_2);w_n\!+\!n;i\vk\cdot\vx\right)
\widetilde{{\cal{O}}}_{i_1i_2\cdots i_n}(\vk) \label{OPE333}
\ee
and  multiply the above expression by $\sigma(\vy)$ (with  $|\vec{y}|\gg|\vx|\!\simeq \!0)$.  Fourier transforming it we get
\be
&&\langle\sigma_{\vec{k}_1}\sigma_{\vec{k}_2}\sigma_{\vec{k}_3}\rangle^\prime\stackrel{k_1\ll k_2}{\sim}
\sum_{n}C_n\int {\rm d}^3 x \,e^{i\vk_2\cdot \vx}\frac{1}{x^{\ell_1+\ell_2}}x^{i_1}x^{i_2}
\cdots x^{i_n}\, \nonumber \\
&&\hspace{3.2cm} \times
\int {\rm d}^3k \,  \Phi\!\left(\!\frac{1}{2}(\ell_1\!-\!\ell_2);w_n\!+\!n;i\vk\cdot\vx\right)
\langle \widetilde{{\cal{O}}}_{i_1i_2\cdots i_n}(\vk)\sigma_{\vk_1}\rangle, \label{s01}
\ee
since $\vk_1$ and $\vk_2$ are the dual vectors to $\vy$ and $\vec{x}$, respectively.
Using again the orthogonality of the two-point function, we get that only the scalar ${\cal{O}}=\sigma$ will contribute to the 
right-hand side of (\ref{s01}), {\it i.e.} $n=0,w_n=w,\ell_2=0,\ell_1=w$. Eq.  (\ref{s01}) turns out to be

\begin{eqnarray}
\langle\sigma_{\vec{k}_1}\sigma_{\vec{k}_2}\sigma_{\vec{k}_3}\rangle^\prime&\stackrel{k_1\ll k_2}{\sim}&
\int {\rm d}^3 x \,e^{i\vk_2\cdot \vx}
\int {\rm d}^3k \,  \Phi\!\left(\!\frac{1}{2}(\ell_1\!-\!\ell_2);w_n\!+\!n;i\vk\cdot\vx\right)
\langle\sigma_{\vk}\sigma_{\vk_1}\rangle\label{s0101}\nonumber\\
&=&
\int {\rm d}^3k 
\int  {\rm d}^3 x \frac{e^{i\vk_2\cdot \vx_1}}{x^{w}}\, (2\pi)^3\,\delta^{(3)}(\vk+\vk_1)\, P_{\vk_1}\,  
\Phi\!\left(\!\frac{w}{2};w;i\vk\cdot\vx\right),
\end{eqnarray}
that is

\be
\fbox{$\displaystyle
\langle\sigma_{\vec{k}_1}\sigma_{\vec{k}_2}\sigma_{\vec{k}_3}\rangle^\prime
\sim (2\pi)^3 P_{\vk_1}\int   
{\rm d}^3 x \frac{e^{i\vk_2\cdot \vx_1}}{x^{w}}\, 
\Phi\!\left(\!\frac{w}{2};w;i\vk_1\cdot\vx\right)+{\rm cyclic}\,\,\,\,\,\,(k_1\ll k_2\sim k_3)$}.
\ee
By expanding the confluent hypergeometric function 
\be
\Phi\!\left(\!\frac{w}{2};w;i\vk_1\cdot\vx\right)\approx 1+\frac{i}{2}\vk_1\cdot\vx-\frac{2+w}{8(1+w)}(\vk_1\cdot\vx)^2+
\cdots
\ee
and using 
\begin{eqnarray}
A_1&=&\int {\rm d}^3 x\, \frac{e^{i\vk_2\cdot \vx_1}}{x^{w}}= \frac{\Gamma(\frac{3-w}{2})\Gamma(w)}{2^w\Gamma(\frac{w}{2})
\Gamma(\frac{3}{2}-w)}
\frac{2}{k_2^w}P_{\vk_2},\nonumber\\
A_2&=&\frac{i}{2}\int {\rm d}^3 x \frac{e^{i\vk_2\cdot \vx_1}}{x^{w}}\vk_1\cdot \vx=
\frac{w\!-\!3}{2}\,  \frac{\vk_1\cdot \vk_2}{k_2^2}A_1,\nonumber\\
A_3&=&\frac{2+w}{8(1+w)}\frac{i}{2}\int {\rm d}^3 x \frac{e^{i\vk_2\cdot \vx_1}}{x^{w}}\vk_1\cdot \vx \nonumber \\
&=&-\alpha_0\, A_1
\left(\frac{k_1^2}{k_2^2}+(w-5)\frac{(\vk_1\cdot\vk_2)^2}{k_2^4}\right)\, , ~~~~~~ \alpha_0=\frac{(2+w)(w+3)}{8(1+w)},
\end{eqnarray}
we finally get the generic form of the three-point correlator in the squeezed limit\footnote{Tree-level computations of the 
three-point correlator (as well as of the four-point correlator) may lead to the presence of 
logarithmic factors $\ln(-k_t\eta)$, where $k_t=(k_1+k_2+k_3)$. For instance, this happens  in the cubic model with interaction ${\cal L} \supset (m/3)\sigma^3$ \cite{falk}. They originate from the fact that the perturbation mode, after horizon crossing, has a nontrivial evolution due to the nonlinearities. Note that the OPE expansion is not sensitive to contact terms so 
factors of the form  $\ln(-k_t\eta)$
cannot be detected in the squeezed limit, However, 
this time dependence is likely to  disappears at the level of correlators of the light fields when  a consistent  resummation  of the IR effects is 
performed \cite{slothriotto}.} 
(up to a model-dependent amplitude)
\be
\fbox{$\displaystyle
\langle\sigma_{\vec{k}_1}\sigma_{\vec{k}_2}\sigma_{\vec{k}_3}\rangle^\prime
\!\sim \!
\frac{(2\pi)^3\Gamma(\frac{3-w}{2})\Gamma(w)}{2^{w-1}\Gamma(\frac{w}{2})\Gamma(\frac{3}{2}\!-\!w)}\frac{1}{k_2^{w}}P_{\vk_1}
P_{\vk_2}\!\left\{\!1\!+\!\alpha_0\left(\!\frac{k_1^2}{k_2^2}\!+\!(w\!-\!5)\frac{(\vk_1\cdot\vk_2)^2}{k_2^4}\right)\right\}\,\,\,\,(k_1\ll k_2\sim k_3)$},\nonumber\\
\ee
\be
\ee
where the cyclic terms  have  been taken into account, reproducing the result (\ref{kk2}).
 Note that terms linear in $|\vk_1|$ from $A_2$ do not contribute as they cancel
 when cyclicity is considered. 
For an almost scale invariant  spectrum $w\approx 0$  the above expression reduces to
\be
\langle\sigma_{\vec{k}_1}\sigma_{\vec{k}_2}\sigma_{\vec{k}_3}\rangle^\prime
\sim 
   P_{\vk_1}P_{\vk_2}\left\{1+\frac{3}{4}\left(\frac{k_1^2}{k_2^2}-5\frac{(\vk_1\cdot\vk_2)^2}{k_2^4}\right)\right\}\,\,\,\,\,\,(k_1\ll k_2\sim k_3).
\ee

\subsection{The four-point function from the OPE and its collapsed limit}
\noindent
We have seen above that the OPE encodes the conformal structure of the conformal field theory. In particular, for primary operators 
the OPE forms a kind of algebra in the sense that the product of two primaries 
at short distance  may be expressed as a series of local operators. For example, the product of two operators
$\sigma^I$ and $\sigma^J$ of conformal dimensions $w_I$ and $w_J$ respectively can be expanded in terms of local operators, collectively
denoted by ${\cal{O}}_{(m)}$ of dimension $w_{{\cal{O}}}$ as  
\be
\sigma^I(\vx)\sigma^J(\vec{0})\approx \sum_{{\cal{O}}} f_{IJ{\cal{O}}}\left\{ C^{(m)}(\vx){\cal{O}}_{(m)}(\vec{0})+\ldots\right\},\label{12}
\ee
where only primaries are needed to be included in the right-hand side of (\ref{12}). 
The coefficient $C^{(m)}(\vx)$ are generally given by
\be
C^{(m)}(\vx)= \frac{x^{i_1}\ldots x^{i_m}}{|\vx|^\ell}\, , ~~~~\ell=w^I+w^J-w_{{\cal{O}}}+m.
\ee
and the dots in (\ref{12}) represent less singular contributions. 
The structure of the OPE (\ref{12}) is similar to a Lie algebra. There, an arbitrary product of generators (the enveloping algebra), can be reduced to a
product of two generators by employing the commutation relation. The same happens here. A generic $n$-point correlator $\langle
\sigma^I(\vx_1)\sigma^J(\vx_2)\cdots\sigma^K(\vx_n)\rangle$ can be reduced to a three-point function, which is specified 
by conformal invariance,  by employing continuously 
the operator product expansion (\ref{12}) \cite{fradkin,fradkin2}. However, such a procedure requires the knowledge of the OPE (\ref{12}) 
and in particular of the coefficients 
$f_{IJ{{\cal{O}}}}$. Of course this is as difficult as the original problem.    

This program can be explicitly seen in the case of the four-point function.  
Contrary to the three-point correlators, four-point correlators are not fully specified by conformal invariance. 
In particular, conformal invariant four-point functions for arbitrary operators ${\cal{O}}^I$ of dimension $w_I$  takes the form 
\be
\hspace{-1cm}&&\langle
{\cal{O}}^I(\vx_1){\cal{O}}^J(\vx_2){\cal{O}}^K(\vx_3){\cal{O}}^L(\vx_4)\rangle=\left(\frac{x_{12}}{x_{14}}\right)^{w_I-w_J}
\left(\frac{x_{14}}{x_{13}}\right)^{w_K-w_L}\frac{g(u,v)}{x_{12}^{\,\,\,w_I+w_J}x_{34}^{\,\,\,w_K+w_L}},
\ee
where  
\be
 u=\frac{x_{12}^2x_{34}^2}{x_{13}^2 x_{24}^2}\, , ~~~~ v=\frac{x_{14}^2x_{23}^2}{x_{13}^2 x^2_{24}} \label{uv}
\ee
are the so-called anharmonic ratios. Therefore, the four-point functions are determined up to an unknown  function $g(u,v)$. Nevertheless, this function has to satisfy certain 
conditions, following basically from the associativity of Eq. (\ref{12}). In fact, we can employ an OPE expansion
along the $(12)(34)$ (or $(14)(23)$) channel. It is easy to see that we get in this case the consistency condition
\be
g(u,v)=\sum_{{\cal{O}}}f_{12{\cal{O}}}f_{34{\cal{O}}}\, G_{w_{\cal{O}},l}(u,v),
\ee
where the sum is over primaries (belonging both to  the (12) and (34) channels) and $G_{w,l}(u,v)$ 
are the so-called conformal blocks. 
We  now restrict ourselves to the case of  scalar operators $\sigma^I$ of equal dimensions $w_I=w$. In such a case,  
the four-point function turns out to be
\be
\hspace{-1cm}&&\langle
\sigma^I(\vx_1)\sigma^J(\vx_2)\sigma^K(\vx_3)\sigma^L(\vx_4)\rangle=
\frac{g^{IJKL}(u,v)}{x_{12}^{\,\,\,2w}x_{34}^{\,\,\,2w}}.       \label{4pt}
\ee
The conformal blocks   $G_{w,l}(u,v)$ have a closed form in $d=2$ and $4$ dimensions \cite{fgg3,petkou1,dolan1,dolan3}. 
However, in $d=3$ their structure is determined by the following  equations  \cite{dolan3}  
\be
&&\frac{(2w -1 ) l}{2l -1} G_{w,l}=
\frac{  (w +l-1)}{2l -1} G_{\Delta ,l-2}+\frac{1}{2} \left\{(w -1 )
   {\cal{F}}_0+\frac{{\cal{F}}_2}{
   l-1}\right\}G_{w +1,l-1}\nonumber \\
 &&  \hspace{2.7cm} -\frac{4w^2  (w -1 )   }{4w^2-1} b\, G_{w +2,l-2},\label{s1}\\
&&\frac{(2w -1 ) l}{2l -1} G_{w ,l}=
\left\{\frac{1}{2}(w +l-3) {\cal{F}}_0 +{\cal{F}}_1\right\}G_{w +1,l-1}
  \nonumber  \\
&& \hspace{2.7cm}
-\frac{4w^2(l-1) }{4w^2-1}b\, G_{\Delta +2,l-2}
    -\frac{2(w +l-1)(l-1) }{2l-1}G_{w
   ,l-2}.\label{s2}
\ee
We have defined 
\be
&&{\cal{F}}_0= \frac{1}{z}+\frac{1}{\bar{z}}-1\,,\qquad
{\cal{F}}_1= (1-z)\partial_z+(1-\bar{z}) \partial_{\bar{z}}\,,\qquad \\
&&
	{\cal{F}}_2= \frac{z-\bar{z}}{z \bar{z}}\left\{z^2(1-z) \partial_z^2-z^2 \partial_z-
\bar{z}^2(1-\bar{z}) \partial_{\bar{z}}^2-\bar{z}^2 \partial_{\bar{z}}\right\},\nonumber \\
&&
b=\frac{(w-l+1)^2}{16(w-l+1)(w-l+2)}
\ee
and $z,\bar{z}$ are given in terms of $u,v$ as 
\be
z=\frac{1}{2}\Big{(}1-v+u+\sqrt{(1-v+u)^2-4u}\Big{)}\, ~~~\bar{z}=\frac{1}{2}\Big{(}1-v+u-\sqrt{(1-v+u)^2-4u}\Big{)},
\ee
or, equivalently,
\be
u=z\bar{z}\, , ~~~v=(1-z)(1-\bar{z}).
\ee
In principle, Eqs. (\ref{s1}) and (\ref{s2}) is a system of equations which can be solved recursively. Although this is 
complicated system,  it simplifies considerably for $z=\bar{z}$. In this case, ${\cal{F}}_2=0$ 
and one may solve either (\ref{s1}) or (\ref{s2}) recursively, which gives \cite{rychkov} 
\be
&&G_{w,0}(z) =
\left(\frac{z^{2}}{1-z}\right)^{w/2}
\, _3F_2\left(\frac{w}{2},\frac{w}{2},\frac{w-1 }{2} ;\frac{w+1 }{2},\frac{2w-1}{2}
   ;\frac{z^2}{4 (z-1)}\right),  
\label{G1}  \\
&&G_{w,1}(z) =\frac{2-z }{2 z} \left(\frac{z^2}{1-z}\right)^{\frac{w +1}{2}}
\, _3F_2\left(\frac{w+1 }{2},\frac{w+1
   }{2},\frac{w }{2} ;\frac{w+2
   }{2},\frac{2w-1}{2} ;\frac{z^2}{4 (z-1)}\right).  \label{G2} 
\ee  
It is noticeable that there is a further simplification in the $z=\bar{z}\to 0$ limit. In this case, 
we get to leading order in $z$
\be
G_{w,l}=c_{w,l}\, z^w +\cdots.
\ee
The coefficients $c_{w,l}$ satisfy the recursion relation 
\be
c_{w,l}=\frac{3(w+l-1)}{(2w-1)l}c_{w,l-2},
\ee
which is solved by
\be
&&c_{w,l}=\left(\frac{3}{2w-1}\right)^{l/2}\frac{\Gamma(\frac{w+l+1}{2})}{\Gamma(\frac{l+2}{2})\Gamma(\frac{w+1}{2})}
\Big{(}1+c_0\big{(}(-)^l-1\big{)}\Big{)}, \nonumber \\
&& c_0=\frac{1}{12}\left\{6-\frac{(3\pi)^{1/2} (2w-1)^{1/2}\Gamma(\frac{1+w}{2})}{\Gamma(1+\frac{w}{2})}\right\}. 
\ee
Since for $z=\bar{z}$ we have $u=z^2$ and $v=(1-z)^2$, we get that  to leading order $g^{IJKL}(u,v)$ is given by
\be
g^{IJKL}(u,v)= g_0^{IJKL}\left(\frac{x_{12}x_{34}}{x_{13} x_{24}}\right)^w+\cdots\,\,\,\,(u\simeq 0,v\simeq 1).
 \ee
Therefore we find that the four-point function in (\ref{4pt})  has the following form in the
$u\simeq 0$ and $v\simeq 1$ limit ($g_0^{IJKL}$ being model-dependent coefficients)
\be
\hspace{-1cm}&&\langle
\sigma^I(\vx_1)\sigma^J(\vx_2)\sigma^K(\vx_3)\sigma^L(\vx_4)\rangle\sim
\frac{g_0^{IJKL}}{x_{12}^{\,\,\,w}x_{13}^{\,\,\,w}x_{24}^{\,\,\,w}x_{34}^{\,\,\,w}}.\label{4ps}
\ee 
By Fourier transforming (\ref{4ps}) we get the four-point function in momentum space 
\be
&&\langle \sigma_{\vk_1}^I\sigma_{\vk_2}^J\sigma_{\vk_3}^K\sigma_{\vk_4}^L\rangle = \int \left(\prod_{i=1}^4 {\rm d}^3\vx_i\right)
e^{i\sum_i \vk_i\cdot\vx_i}
\langle
\sigma^I(\vx_1)\sigma^J(\vx_2)\sigma^K(\vx_3)\sigma^L(\vx_4)\rangle.
\ee
Using Eqs. (\ref{xw}) and (\ref{4ps}) we get 
\be
&&\langle \sigma_{\vk_1}^I\sigma_{\vk_2}^J\sigma_{\vk_3}^K\sigma_{\vk_4}^L\rangle \sim 
\left(\frac{\Gamma(\frac{3-w}{2})}{2^w\pi^{3/2}
\Gamma(\frac{w}{2})}\right)^4\int \left(\prod_{i=1}^4 {\rm d}^3\vx_i\right)\left(\prod_{i=1}^4 {\rm d}^3\vq_i\right) 
\nonumber \\
&&\hspace{2.5cm}\times e^{i\sum_i \vk_i\cdot\vx_i} \frac{e^{-i\vq_1\cdot \vx_{12}}e^{-i\vq_2\cdot \vx_{24}}e^{-i\vq_3\cdot 
\vx_{31}}e^{-i\vq_4\cdot \vx_{43}}}{|\vq_1|^{3-w}
|\vq_2|^{3-w}v|\vq_3|^{3-w}|\vq_4|^{3-w}}. \label{4xl}
\ee
It is clear that the internal momenta $\vq_i$ are the eigenvalues of the operators
\be
\vq_1=-i\vec{\partial}_{12}, ~~~
\vq_2=-i\vec{\partial}_{31}, ~~~\vq_3=-i\vec{\partial}_{24}, ~~~\vq_4=-i\vec{\partial}_{43}. \label{qq}
\ee
Performing  the $\vx_i$ integrations in (\ref{4xl}), we get $\delta$-functions which specify the external momenta $\vk_i$ as
\be
 \vk_1=\vq_1-\vq_3, ~~~\vk_2=\vq_2-\vq_1, ~~~\vk_3=\vq_3-\vq_4, ~~~\vk_4=\vq_4-\vq_2. \label{kq}
\ee
Now, what does the limit $u\simeq 0$ and $v\simeq 1$ correspond to in terms of space distances? 
Using Eqs. (\ref{qq}) and (\ref{kq}) we may find the limit $u\simeq  0$ and $v\simeq 1$ in momentum space; then  
from (\ref{uv}) we get that these limits correspond to
\be
x_{12}x_{34}\ll x_{13} x_{24} ~~~~{\rm and}~~~~x_{14}x_{23}\sim x_{13}x_{24} \label{x12}.
\ee
We can satisfy the relations (\ref{x12})  by taking 
\be
x_{12}\ll {\rm{min}}(x_{13},  x_{24})  ~~~~{\rm and}~~~~ x_{34}\ll {\rm{min}}( x_{13},x_{34}).   
\ee
For the internal momenta $\vq$ this implies  that 
\be
|\vq_1|\gg  {\rm{max}}(|\vq_2|,  |\vq_3|)  ~~~~{\rm and}~~~~|\vq_4|\gg  {\rm{max}}(|\vq_2|,  |\vq_3|),
\ee
which, for the external momenta, implies
\be
|\vk_{12}|=|\vq_2-\vq_3|\ll {\rm{min}}(|\vk_i|) ~~~(i=1,\cdots,4).
\ee
This particular configuration is indicated in Fig. 2(a)  in  real space and in Fig. 3(a) in momentum space, respectively.  It is known in the literature as the collapsed configuration. Therefore we conclude that the generic NG four-point correlator in the collapsed configuration is of the form
(as  $x_{13}\approx x_{24}$)
\be
\fbox{$\displaystyle
\langle
\sigma^I(\vx_1)\sigma^J(\vx_2)\sigma^K(\vx_3)\sigma^L(\vx_4)\rangle=
\frac{g_{0}}{x_{12}^{\,\,\,w}x_{13}^{\,\,\,2 w}x_{34}^{\,\,\,w}}$}.\label{4pss}
\ee 
By Fourier transforming (using again (\ref{xw})) we find that the generic NG four-point correlator in momentum space in the collapsed limit is (up to a model-dependent coefficient) 
\be
\langle \sigma_{\vk_1}^I\sigma_{\vk_2}^J\sigma_{\vk_3}^K\sigma_{\vk_4}^L\rangle^\prime \sim
 \frac{1}{|\vk_{12}|^{3-2w}|\vk_2|^{3-w} |\vk_4|^{3-w}}+\,\,{\rm permutations}\,\,\,\,\,\,\,\,(|\vk_{12}|\to 0),
 \ee
 or
 \be
 \fbox{$\displaystyle
 \langle \sigma_{\vk_1}^I\sigma_{\vk_2}^J\sigma_{\vk_3}^K\sigma_{\vk_4}^L\rangle^\prime \sim
 |\vk_2|^{-w}|\vk_4|^{-w}P_{|\vk_{12}|}P_{\vk_2}P_{\vk_4}+\,\,{\rm permutations}\,\,\,\,\,\,\,(|\vk_{12}|\to 0)$}.
\ee
Before closing this section, we notice that there is another possibility to realize the condition 
(\ref{x12}), namely we can consider the configuration 
\begin{figure}[t!]
\begin{center}
 \includegraphics[scale=0.5]{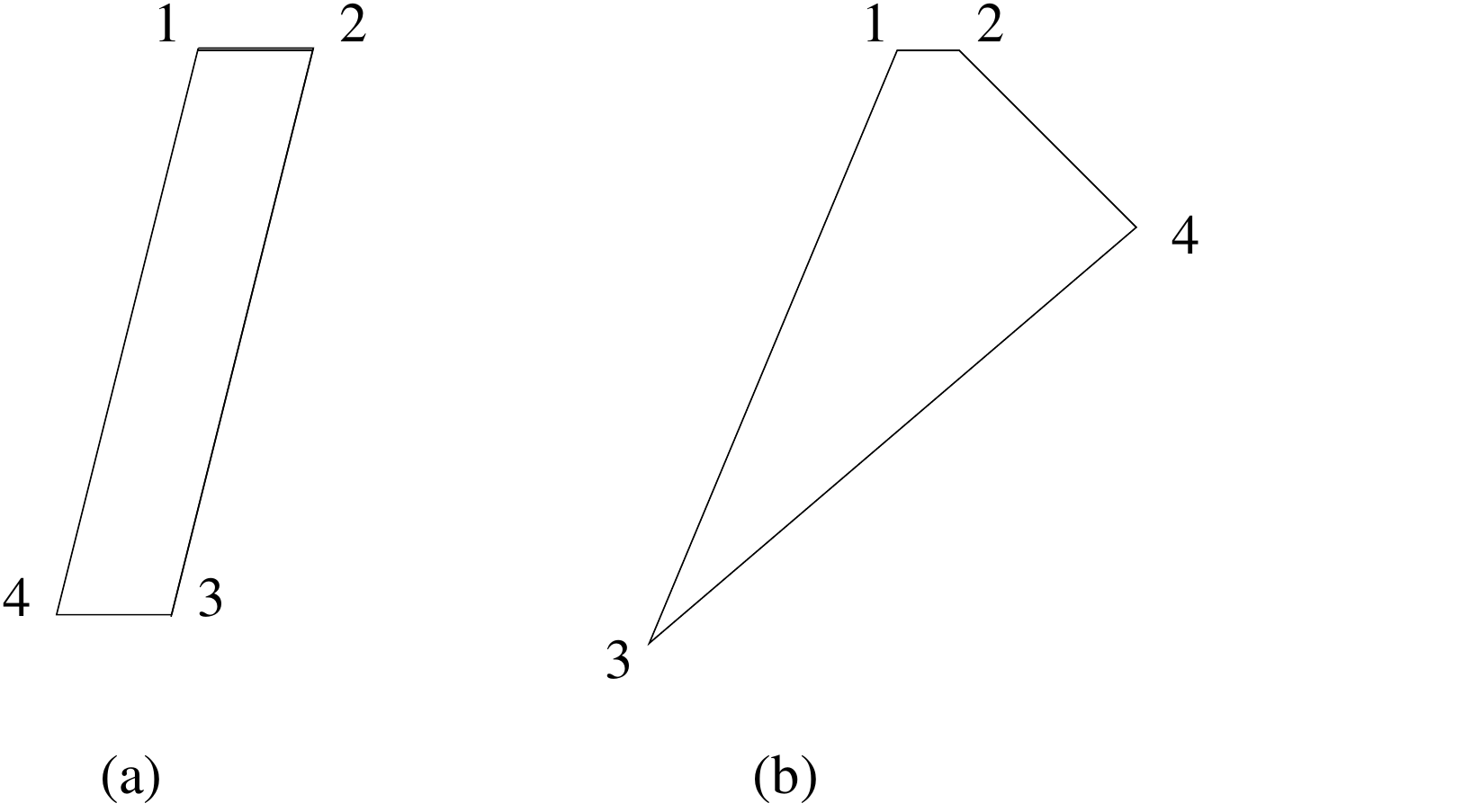}
\caption{(a) Collapsed configuration projected on a plane in space where  $x_{12}\approx x_{34}\approx 0$ 
with $x_{13}\gg x_{12},x_{34}$. (b) Double squeezed 
configuration where $x_{34}\approx x_{13}\gg x_{24}\gg x_{12}\approx 0$. }
\vskip .5in
 \includegraphics[scale=0.5]{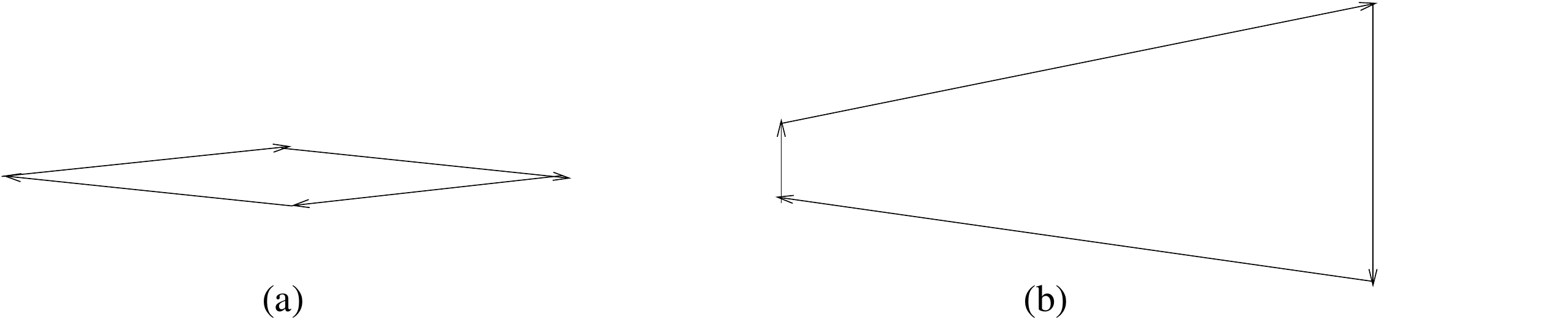}
\caption{(a) Collapsed and  (b) double squeezed shapes in momentum space. }
\end{center}
\end{figure}
\be
x_{34}\approx x_{13}\gg x_{24}\gg x_{12}\simeq 0.
\ee
The internal momenta $\vq_i$ then obey
\be
|\vq_2|\approx |\vq_4|\ll |\vq_3|\ll |\vq_1|
\ee
and correspondingly the external momenta $\vk_i$ satisfy the relation
\be
|\vk_1|\approx |\vk_2|\gg |\vk_3|\gg |\vk_4|.
\ee
This double squeezed configuration is drawn  in Fig.  2(b) in real space  and in Fig. 3(b) in momentum space. Since $x_{13}\simeq x_{34}$,  we 
get for the four-point function

\be
\hspace{-1cm}&&\langle
\sigma^I(\vx_1)\sigma^J(\vx_2)\sigma^K(\vx_3)\sigma^L(\vx_4)\rangle\sim
\frac{g_0}{x_{12}^{\,\,\,w}x_{24}^{\,\,\, w}x_{34}^{\,\,\,2w}}.\label{4psss}
\ee 
By Fourier transforming and  employing   Eq. (\ref{xw}) we find  
\be
\langle \sigma_{\vk_1}^I\sigma_{\vk_2}^J\sigma_{\vk_3}^K\sigma_{\vk_4}^L\rangle^\prime& \sim&  
 \frac{1}{|\vk_1|^{3-w}|\vk_2|^{3-2w} |\vk_4|^{3-w}}=\nonumber \\
&=& |\vk_1|^{-w}|\vk_4|^{-w}P_{\vk_1}P_{\vk_3}P_{\vk_4} 
~~~~~(|\vk_1|\approx |\vk_2|\gg |\vk_4|\gg |\vk_3|). 
\ee

\section{On the Suyama-Yamaguchi inequality}
The collapsed limit of the four-point correlator is particularly important because, together 
with the squeezed limit of the three-point correlator, it may lead to the so-called Suyama-Yamaguchi (SY) inequality \cite{SY}. Consider  a class of
multi-field models which satisfy the following conditions: a) 
scalar fields are responsible for generating curvature
perturbations and b) the fluctuations in scalar fields at the horizon crossing
are scale invariant and gaussian. The second condition amounts to assuming that the connected three- and four-point correlations of the $\sigma^I$ fields vanish and that the NG is generated
at superhorizon scales. If so, the three- and four-point correlators of the comoving curvature perturbation (\ref{zeta3}) and  (\ref{zeta4}) respectively reduce to
\be
B_\zeta(\vk_1,\vk_2,\vk_3)= N_I N_{JK}N_{L}\left(P^{IK}_{\vec{k}_1}P^{JL}_{\vec{k}_2}+2\,\,{\rm permutations} \label{zeta3reduced}
\right)
\ee
and
\begin{eqnarray}
T_\zeta(\vk_1,\vk_2,\vk_3,\vk_4)&=&
N_{IJ} N_{KL}N_{M}N_N\left(P^{JL}_{\vec{k}_{12}}P^{IM}_{\vec{k}_{1}}P^{KN}_{\vec{k}_{3}}
+11\,\,{\rm permutations}
\right)
\nonumber\\
&+&N_{IJK} N_{L}N_{M}N_N\left(P^{IL}_{\vec{k}_{1}}P^{JM}_{\vec{k}_{2}}P^{KN}_{\vec{k}_{3}}
+3\,\,{\rm permutations}
\right),
 \label{zeta4reduced}
\end{eqnarray}
Notice in particular that in the collapsed limit the last term of the four-point correlator (\ref{zeta4reduced})  
is subleading. By defining the nonlinear parameters $f_{\rm NL}$ and $\tau_{\rm NL}$
as
\begin{eqnarray}
f_{\rm NL}&=&\frac{5}{12}\frac{\langle \zeta_{\vk_1}\zeta_{\vk_2}
\zeta_{\vk_3}\rangle^\prime}{P^\zeta_{\vk_1}P^\zeta_{\vk_2}}\,\,\,\,\,\,\,\,(k_1\ll k_2\sim k_3),\nonumber\\
\tau_{\rm NL}&=&\frac{1}{4}\frac{\langle \zeta_{\vk_1}\zeta_{\vk_2}\zeta_{\vk_3}\zeta_{\vk_4}\rangle^\prime}{P^\zeta_{\vk_1}
P^\zeta_{\vk_3}P^\zeta_{\vk_{12}}}\,\,\,\,\,\,\,\,(\vk_{12}\simeq  0), \label{tf2}
\end{eqnarray}
and making use of the Cauchy-Schwarz inequality one can prove the SY inequality (at the tree-level) $\tau_{\rm NL}\geq (6 f_{\rm NL}/5)^2$, where the equality holds in the case of a single scalar field \cite{SY}. 
%

A crucial question is if the SY inequality will still hold if the  NG correlators of the fields $\sigma^I$ do not vanish. Indeed, the conformal symmetry imposes that the squeezed limit of the three-point correlator as well as the collapsed limit of the four-point correlator have the same shapes of those present in Eqs. (\ref{zeta3reduced}) and Eqs. (\ref{zeta4reduced}), respectively. 
Therefore, one might expect a contamination of the inequality if the light scalar fields are NG at horizon crossing. 
Since being NG at horizon crossing requires simply that the light fields have self-interactions, a contamination of 
the SY inequality is rather plausible. However, the SY inequality is still valid even in the case of 
intrinsically NG fields.  A step towards this proof was taken in Refs.  \cite{SY1} were the generic case of NG fields was considered. Nevertheless, it was assumed there that the coefficients
$f_{\rm NL}$ and $\tau_{\rm NL}$ in Eq. (\ref{tf2}) were momentum-independent, see {\it e.g.} the discussion between Eqs. (2.13) and (2.14) of Assassi et al. \cite{SY1}. 

To explicitly demonstrate the SY inequality for NG fields, let us consider the OPE expansions for the two fields $\sigma^I$  
and $\sigma^J$    in the (12) channel at the coincident point 
\be
&&\sigma^I(\vx_1)\sigma^J(\vx_2)\!=\!\left(\frac{C_0^{IJ}(w)}{x_{12}^{2w}}+\frac{{C^{IJ}}_M(w)}{x_{12}^w}\sigma^M(\vx_2)+\cdots\!\right)\!
= \nonumber \\
&&\hspace{2.3cm}
=\sum_{n,s}\frac{{C_{ns}^{IJ}}_M(w)}{x_{12}^{2w+w_n+s}}x_{12}^{i_1}x_{12}^{i_2}\cdots 
x_{12}^{i_s}{\cal{O}}^{M(ns)}_{i_1i_2\cdots i_s}(\vx_2) ~~~~(x_{12}\simeq  0) 
\nonumber\\
\label{122}
\ee
and similarly in the (34) channel at the coincident point
\be
&&\sigma^K(\vx_3)\sigma^L(\vx_4)\!=\!\left(\frac{C_0^{KL}(w)}{x_{34}^{2w}}+\frac{{C^{KL}}_M(w)}{x_{34}^w}\sigma^M(\vx_4)\!+\!\cdots\!\right)
= \nonumber \\
&&\hspace{2.4cm}
=\sum_{n,s}\!\frac{{C_{ns}^{KL}}_M(w)}{x_{34}^{2w+w_n+s}}x_{34}^{i_1}x_{34}^{i_2}\cdots 
x_{34}^{i_s}{\cal{O}}^{M(ns)}_{i_1i_2\cdots i_s}(\vx_4)~~~~(x_{34}\simeq 0).\nonumber\\
 \label{34}
\ee
The four-point function in the collapsed limit 
\be
\langle\sigma^I(\vx_1)\sigma^J(\vx_2)\sigma^K(\vx_3)\sigma^L(\vx_4)\rangle ~~~(x_{12}\simeq 0\,\,{\rm and \,\,} x_{34}\simeq  0)
\ee 
can be expressed, using  the OPE's  (\ref{12}) and (\ref{34}),  as
\begin{eqnarray}
\langle\sigma^I(\vx_1)\sigma^J(\vx_2)\sigma^K(\vx_3)\sigma^L(\vx_4)\rangle
&=&\Big< \left(\sum_{n,s}\frac{{C_{ns}^{IJ}}_M}{x_{12}^{2w+w_n+s}}x_{12}^{i_1}x_{12}^{i_2}\cdots 
x_{12}^{i_s}{\cal{O}}^{M(ns)}_{i_1i_2\cdots i_s}(\vx_2)\right)
\nonumber \\
&\times & \left(\sum_{n',s'}\frac{{C_{ns}^{KL}}_N}{x_{34}^{2w+w_n+s}}x_{34}^{i_1}x_{34}^{i_2}\cdots 
x_{34}^{i_s}{\cal{O}}^{N(n's')}_{i_1i_2\cdots i_s'}(\vec{x}_4)\right)\Big>.
\end{eqnarray}
Due to the orthogonality of the two point function
\be
\langle{\cal{O}}^{M(ns)}_{i_1i_2\cdots i_s}(\vx_2){\cal{O}}^{N(n's')}_{j_1j_2\cdots j_s'}(\vec{x}_4)
\rangle=\frac{C_{(ns)}^{MN} t_{i_1\cdots i_s j_1\cdots j_s}}{x_{24}^{\,\,\, 2w_n}}\delta_{ss'}\delta_{nn'},
\ee
where $ t_{i_1\cdots i_s j_1\cdots j_s}$ is a positive tensor build up from Kronecker deltas, 
the four-point function in the collapsed limit  may be expressed as
\be
\langle\sigma^I(\vx_1)\sigma^J(\vx_2)\sigma^K(\vx_3)\sigma^L(\vx_4)\rangle=\frac{C_0^{IJA}{C_0^{KL}}_A}{x_{12}^{2w}x_{34}^{2w}}+
\frac{{C^{IJ}}_A{C^{KL}}_B}{x_{12}^wx_{34}^w}\langle\sigma^A(\vx_2)\sigma^B(\vx_4)\rangle+\cdots. \label{ab}
\ee
Denoting by $\langle x_{12}|t|x_{34}\rangle=x_{12}^{i_1}x_{12}^{i_2}\cdots 
x_{12}^{i_s} \, t_{i_1\cdots i_s j_1\cdots j_s}\, x_{34}^{j_1}x_{34}^{j_2}\cdots 
x_{34}^{j_s}$ and $C^A=N_IN_JC^{IJA}$,  we can write 
\begin{eqnarray}
N_IN_JN_KN_L\langle\sigma^I(\vx_1)\sigma^J(\vx_2)\sigma^K(\vx_3)\sigma^L(\vx_4)\rangle&=&
\frac{C_0^{A}C_{0A}}{x_{12}^{2w}x_{34}^{2w}}+
\frac{{C}_A{C}_B}{x_{12}^wx_{34}^w}\langle\sigma^A(\vx_2)\sigma^B(\vx_4)\rangle+\cdots  \nonumber \\
&=&\sum_{n,s}
\frac{{C_{(ns)}^A}{C_{(ns)A}}\langle x_{12}|t|x_{34}\rangle}{x^{2w+w_n+s}} \geq 0, \label{pos}
\ee
from which we deduce that 
\be
N_IN_JN_KN_L\langle\sigma^I(\vx_1)\sigma^J(\vx_2)\sigma^K(\vx_3)\sigma^L(\vx_4)\rangle \ge
\frac{{C}_A{C}_B}{x_{12}^wx_{34}^w}\langle\sigma^A(\vx_2)\sigma^B(\vx_4)\rangle
~~~(x_{12}\simeq 0\,\,{\rm and \,\,} x_{34}\simeq  0).
 \label{NNN}
\ee
On the other side, the two-point correlator is (up to an irrelevant overall normalization factor)
\be
\langle\sigma^I(\vx_2)\sigma^J(\vx_4)\rangle=\frac{1}{x_{24}^{2w}}\delta^{IJ}, \label{2cor}
\ee
so that the expression (\ref{NNN}) may be explicitly written as 
\be
N_IN_JN_KN_L\langle\sigma^I(\vx_1)\sigma^J(\vx_2)\sigma^K(\vx_3)\sigma^L(\vx_4)\rangle \ge
  \frac{C_AC^A}{x_{12}^wx_{34}^wx_{24}^{2w}}. \label{svx}
\ee
By Fourier transforming (\ref{2cor}) we get 
\be 
\langle \sigma^I_{\vk_1} \sigma^J_{\vk_2}\rangle^\prime=\frac{B(2w)}{k^{3-2w}}\delta^{IJ}
\ee
where 
\be
B(w)=2^{3-w}\pi^{\frac{3}{2}}\frac{\Gamma\left(\frac{3-w}{2}\right)}{\Gamma\left(\frac{w}{2}\right)}.
\ee
Therefore we find that the power spectrum of the comoving curvature perturbation is
\be
\langle \zeta_{\vk_1}\zeta_{\vk_2}\rangle^\prime=P^\zeta_{\vk_1}=
N_I N_J P^{IJ}_{\vk_1}=\frac{N_I N^I B(2w)}{k_1^{3-2w}}.
\ee
By Fourier transforming both sides of (\ref{svx}) we get
\be
&&N_I N_J N_K N_L T^{IJKL}_{\vec{k}_1\vec{k}_2\vec{k}_3\vec{k}_4}=
N_I N_J N_K N_L\langle\sigma^I_{\vk_1}\sigma^J_{\vk_2}\sigma^k_{\vk_3}\sigma^L_{\vk_4}\rangle' \nonumber \\
&&\ge \frac{C_AC^AB^2(w)B(2w)}{|\vk_1|^{3-w}|\vk_3|^{3-w}|\vk_{12}|^{3-2w}}=\frac{C_AC^A}{(N_IN^I)^3}P^\zeta_{\vk_{12}}P^\zeta_{\vk_1}
P ^\zeta_{\vk_3}\gamma(k_1)\gamma(k_3), \label{svx1}
\ee
where 
\be
\gamma(k)=\frac{B(w)}{B(2w)}k^{-w}.
\ee
Similarly, 
the three-point correlator in the squeezed limit  can be evaluated 
by employing the OPE (\ref{12}) as
\be
\langle\sigma^I(\vx_1)\sigma^J(\vx_2)\sigma^K(\vx_3)\rangle=
\Big<\left(\frac{C^{IJ}_0}{x_{12}^{2w}}+\frac{C^{IJ}_A}{x_{12}^w}\sigma^A(\vx_2)+\cdots\right)
\sigma^K(\vx_3)\Big>.
\ee
Using again the orthogonality of the two-point functions,  only the correlator (\ref{2cor})
will contribute to the sum above, so that 
\be
\langle\sigma^I(\vx_1)\sigma^J(\vx_2)\sigma^K(\vx_3)\rangle=\frac{{C^{IJ}}_A}{x_{12}^w}\langle\sigma^A(\vx_2)\sigma^K(\vx_3)\rangle
=\frac{C^{IJK}}{x_{12}^wx_{23}^{2w}}  ~~~(x_{12}\simeq 0). \label{3xv}
\ee
Again Fourier transforming (\ref{3xv}) we get 
\be
B^{IJK}_{\vk_1\vk_2\vk_3}=
\frac{C^{IJK}B(w) B(2w)}{|\vk_3|^{3-w}|\vk_{1}|^{3-2w}}=\frac{C^{IJK}B(w) B(2w)}{(N_AN^A)^2} 
P^\zeta_{\vk_{1}}P^\zeta_{\vk_3}\gamma(k_3).
\ee
Using  Eq. (\ref{zeta4}) we get   in the collapsed limit for the four-point function and the squeezed limit for the three-point function 
\be 
4\tau_{\rm NL}=\frac{T_\zeta(\vk_1,\vk_2,\vk_3,\vk_4)}{P^\zeta_{\vk_{12}}P^\zeta_{\vk_1}P^\zeta_{\vk_3}}&=&
\frac{N_I N_J N_K N_L }{P^\zeta_{\vk_{12}}P^\zeta_{\vk_1}P^\zeta_{\vk_3}}T^{IJKL}_{\vec{k}_1\vec{k}_2\vec{k}_3\vec{k}_4}\nonumber\\
&+& \frac{N_{IJ} N_{K}N_{L}N_M}{P^\zeta_{\vk_{12}}P^\zeta_{\vk_1}P^\zeta_{\vk_3}}\left(P^{IK}_{\vec{k}_1}B^{JLM}_{\vec{k}_{12}\vec{k}_3\vec{k}_4}+11\,\,{\rm permutations}
\right)\nonumber\\
&+&\frac{N_{IJ} N_{KL}N_{M}N_N}{P^\zeta_{\vk_{12}}P^\zeta_{\vk_1}P^\zeta_{\vk_3}}\left(P^{JL}_{\vec{k}_{12}}P^{IM}_{\vec{k}_{1}}P^{KN}_{\vec{k}_{3}}
+11\,\,{\rm permutations}
\right)
\nonumber\\
&+&\frac{N_{IJK} N_{L}N_{M}N_N}{P^\zeta_{\vk_{12}}P^\zeta_{\vk_1}P^\zeta_{\vk_3}}\left(P^{IL}_{\vec{k}_{1}}P^{JM}_{\vec{k}_{2}}P^{KN}_{\vec{k}_{3}}
+3\,\,  {\rm permutations}.
\right) 
 \label{zeta44}
\end{eqnarray}
In the squeezed limit  there are four leading terms in  the second and four relevant terms in  the third line of (\ref{zeta44}) while the last line in (\ref{zeta44}) is subleading. Using now Eqs. (\ref{svx1}) and (\ref{zeta44}), 
we may deduce  the inequality 
\be
  4\tau_{\rm NL}\ge \gamma(k_1)\gamma(k_3) \frac{C_AC^A}{(N_AN^A)^3}+4\gamma(k_1)\frac{C^IN_{IJ}N^L}{(N_IN^I)^3}+
4\frac{N^IN_{JI}N^{JK}N_K}{(N_IN^I)^3}.
\ee
It can easily be checked that the right-hand side of the above inequality can be written as  ($\gamma(k_1)\approx \gamma(k_3)$)
\be
\gamma(k_1)\gamma(k_3) \frac{C_AC^A}{(N_AN^A)^3}+4\gamma(k_1)\frac{C^IN_{IJ}N^L}{(N_IN^I)^3}+
4\frac{N^IN_{JI}N^{JK}N_K}{(N_IN^I)^3}=4 \frac{D_AD^A}{(N_IN^I)^3}
\ee
where 
\be
D_A=\frac{1}{2}\gamma(k_3) C_I+N_{IJ}N^J.
\ee
Thus we have that 
\be
\tau_{\rm NL}\ge \frac{D_AD^A}{(N_IN^I)^3}. \label{tf}
\ee
Similarly, using Eq. (\ref{zeta3}) we find that $f_{\rm NL}$ is given in the squeezed limit $k_1\ll k_2\sim k_3$ 
\be
\frac{12}{5}f_{\rm NL}=
\frac{B_\zeta(\vk_1,\vk_2,\vk_3)}{P^\zeta_{\vk_1}P^\zeta_{\vk_3}}=
\frac{N_I N_J N_K}{P^\zeta_{\vk_1}P^\zeta_{\vk_3}} B^{IJK}_{\vec{k}_1\vec{k}_2\vec{k}_3}+ 
\frac{N_I N_{JK}N_{J}}{P^\zeta_{\vk_1}P^\zeta_{\vk_3}}
\left(P^{IK}_{\vec{k}_1}P^{JL}_{\vec{k}_3}+2\,\,{\rm permutations}\right),\label{zeta33}
\ee
from which we find  
\be
\frac{6}{5}f_{\rm NL}=\frac{1}{2}\gamma(k_3)\frac{C^IN_I}{(N_I N^I)^2}+\frac{N^IN_{IJ}N^J}{(N_IN^J)^2}=
\frac{D_IN^I}{(N_I N^I)^2}. \label{ft}
\ee
Then, applying  the Cauchy-Schwarz inequality 
\be
(D_IN^I)^2\leq D_ID^IN_JN^J
\ee
to  Eqs. (\ref{tf}) and (\ref{ft}) we finally get  
\be
\fbox{$\displaystyle
\tau_{\rm NL}\ge \left(\frac{6}{5}f_{\rm NL}\right)^2$}\,\,\,\,\,({\rm also}\,\,{\rm for}\,\,{\rm NG}\,\,{\rm fields}). \label{SYin}
\ee
 This completes the proof that the  SY inequality  in all multifield models where the NG comes from 
 light scalar fields other than the inflaton even when such 
 light scalar fields are NG at horizon crossing. Loop corrections from  the universal superhorizon NG
 part of the comoving curvature perturbation were shown not to change the inequality \cite{sloop}. 
Indeed, in an exact conformal invariant theory, this result would be
 robust against loop corrections as the arguments on  NG correlators of the light fields we worked out through the OPE's are  
 valid at any order of perturbation theory. 

\subsection{A further generalization of the Suyama-Yamaguchi inequality}
\noindent
Before closing this section we would like to discuss two issues:   how much the SY inequality  depends on our assumption that the system enjoys the conformal symmetry and 
 what modifications are introduced due to quantum effects. 
 
 Concerning the first point,  the  conformal symmetry is not 
really crucial. 
In fact, assuming that Wilsonian OPE holds, we expect the short distance bahaviour of a set of fields $\sigma^I(\vx)$ to be
\be
\sigma^I(\vx)\sigma^J(\vec{y})\stackrel{\vx\to\vec{y}}{\sim}\sum_{n} C_{n}(\vx-\vy)
{\cal{O}}_{n}(\vec{y})\label{OPE22}.
\ee
On pure dimensional grounds and naive dimensional counting, the coefficients $C_n$ should behave like 
\be
C_n\sim\left(\frac{1}{|\vx-\vy|}\right)^{w_I+w_J-w_n} \label{c_i},
\ee
where $w_I,w_J,w_n$ are the dimensions of the $\sigma^I,\sigma^J$ and ${\cal{O}}_n$ operators, respectively. 
What is relevant is to 
specify the most singular term in this expansion. Clearly, the highest the dimension of the operators ${\cal{O}}_n$ 
the less singular
the coefficient $C_n$ \cite{Wilson,wilson,zim}. Thus, we will have for example
\be
\sigma^I(\vx_1)\sigma^J(\vx_2)\!=\!\left(\frac{C_0^{IJ}}{x_{12}^{w_I+w_J}}+\frac{{C^{IJ}}_M}{x_{12}^{w_I+w_J-w_M}}
\sigma^M(\vx_2)+\cdots\!\right),\!  \label{cc}
\ee
where the dots stands for less singular terms. Doing the same for $\sigma^K(\vx_3)\sigma^L(\vx_4)$, we find that the 
four-point function, in the $x_{12},x_{34}\to 0$ limit, is given again as 
\begin{eqnarray}
N_I N_J N_K N_L T^{IJKL}_{\vec{k}_1\vec{k}_2\vec{k}_3\vec{k}_4}&=&
N_I N_J N_K N_L\langle\sigma^I_{\vk_1}\sigma^J_{\vk_2}\sigma^k_{\vk_3}\sigma^L_{\vk_4}\rangle' 
\ge \frac{C_AC^B C^{AB} B(w_1)B(w_2)B(2w)}{|\vk_1|^{3-w_1}|\vk_3|^{3-w_2}|\vk_{12}|^{3-2w}}
\nonumber\\
&=&\frac{C_AC_B C^{AB}}
{(N_IN_J C^{IJ})^3}P^\zeta_{\vk_{12}}P^\zeta_{\vk_1}
P ^\zeta_{\vk_3}\gamma(k_1)\gamma(k_3), \label{4p1}
\end{eqnarray}
where $w_1=w_I+w_J-w_A, ~w_2=w_K+w_L-w_B,~2w=w_A+w_B$ and (up to a normalization constant)
\be
P^\zeta_{\vk_1}=
\frac{N_I N_J C^{IJ} B(w_I+w_J)}{k_1^{3-{w_I+w_J}}}.
\ee
The coefficients $C^{IJ}$ are  defined through 
\be 
\langle \sigma^I_{\vk_1} \sigma^J_{\vk_2}\rangle^\prime=C^{IJ}\frac{B(2w)}{k_1^{3-2w}} \label{ccc}
\ee
and 
\be
\gamma(k_1)=\frac{B(w_1)}{B(2w_1)}k_1^{-w_1}, ~~~
\gamma(k_3)=\frac{B(w_2)}{B(2w_2)}k_3^{-w_2}.
\ee
Similarly,  we find for the bispectrum the dominant contribution
\be
B^{IJK}_{\vk_1\vk_2\vk_3}=
\frac{{C^{IJ}}_AC^{AK}B(w_1) B(2w)}{|\vk_3|^{3-w_1}|\vk_{1}|^{3-2w}}=\frac{{C^{IJ}}_AC^{AK}B(w_1) B(2w)}{(N_AN_B C^{AB})^2} 
P^\zeta_{\vk_{1}}P^\zeta_{\vk_3}\gamma(k_3).
\ee
Then, following the same steps from Eq. (\ref{zeta44}) to Eq. (\ref{ft}), the SY condition (\ref{SYin}) follows when the inequality
\be
(D_IN_J C^{IJ})^2\leq D_ID_J C^{IJ}N_KN_L C^{KL} \label{cn}
\ee
is implemented. The above discussion avoids any reference to  conformal invariance. It is based simply on the 
short-distance expansion of the product of two-operators in  Eq. (\ref{cc}). Although, we could not  use  the orthogonality of the 
two-point function for operators of different dimensions as we did in the conformal case, based on the fact that 
we are interested in the collapsed limit, 
we have kept only the  dominant most singular term. All the other terms are subleading  and therefore 
do not contribute in the collapsed limit.

This discussion is valid for any fields with arbritary dimension.  Clearly, the   inflaton field can be one of these 
fields and still the SY inequality is valid for this case as well. Note that we could not  come 
to this conclusion previously in the conformal case, as
a time-evolving inflaton background breaks the special conformal  scale symmetry. However, 
here we can deduce that SY inequality holds also when the inflation field plays a role in determining the cosmological perturbations. The
SY inequality is  more a consequence of fundamental  physical principles 
rather than of pure mathematical arrangements. For example, the  inequality (\ref{cn}) is true only for a positive definite
matrix $C^{IJ}$. A negative definite $C^{IJ}$ would lead to violation of the inequality, but this would require,  for example,   ghost-like scalars 
among the light  $\sigma^I$ fields.   The observation of a strong  violation of the inequality will then have profound 
implications for inflationary models 
as it will imply either that multifield inflation cannot be responsible for
generating the observed fluctuations independently of the details of the model or that some new non-trivial degrees of freedom play a role during inflation. We will give an example of such a case in the next section.

The second point concerns quantum effects. Clearly, although at the tree-level the behaviour of the coefficients
$C_n$'s is determined by the relation  (\ref{c_i}), the 
 renormalization effects will modify it. In fact,   the scaling 
properties of the $C_n$'s  will be  given by the Callan-Symanzik equation  with a  particular operator
 mixing. For example,  for  asymptotically free theories, deviations from canonical scaling is characterized by  multiplicative 
logarithmic functions. As a result, the quantum effects change the functional form of the coefficient $C_n$' s in  the 
short distance expansion of the operators (\ref{cc}). This change might show up in the thee- and four-point function in the 
collapsed limit 
as  momentum-dependent $f_{\rm NL}$ and $\tau_{\rm NL}$. Therefore, renormalization will induce
 in general a different functional  momentum-dependence of  $f_{\rm NL}$ and $\tau_{\rm NL}$,  which might lead to violation of the SY inequality 
for certain range of momenta. Of course, this is model-dependent problem and should be analyzed case by case.

\section{Logarithmic Conformal Field Theories}
\noindent
There is another class of conformal  theories, namely the logarithmic CFT's \cite{Gurarie}, 
which can be of interest from the cosmological point of view for two reasons. First the perturbations
in exact de Sitter are  not scale invariant even in the limit of zero mass. Secondly, it provides one example in which the SY is reversed.

These are theories characterized by the appearance
of logarithms in correlation functions  due to logarithmic short-distance singularities in the OPE. These
singularities are connected to special operators  having conformal dimensions degenerate with those of the usual primary
operators. It is this degeneracy that is at the origin of the appearance of the  logarithms \cite{Gurarie}.

To be more concrete, let us consider fields $\Phi$ and $\Psi$ on de Sitter background with action

\be
S_2=\int {\rm d}^4 x\sqrt{-g}\left(-\partial_\mu\Phi \partial^\mu\Psi-m^2\Phi \Psi-\frac{\mu^2}{2} \Psi^2\right). \label{slog}
\ee
The equations of motions are simply
\be
&&\Box \Psi-m^2 \Psi=0, \label{psi}\\
&& \Box \Phi-m^2 \Phi-\mu^2\Psi=0. \label{phi}
\ee
As usual, the conformal dimensions can be calculated by the asymptotic form of the  space-independent 
 solution $\Phi(\eta)$ and $\Psi(\eta)$ of Eqs. (\ref{psi}) and (\ref{phi}), which are easily found to be
\be
\Psi\sim \eta^w, ~~~\Phi\sim \eta^w \log \eta, ~~~~ w=\frac{3}{2}\left(1-\sqrt{1-\frac{4m^2}{9H^2}}\right).
\ee
Clearly, the scaling of $\Psi$ and $\Phi$ is not conventional as they transform under rescalings as
\be
\Psi\to \lambda^w \Psi, ~~~\Phi\to \lambda^w(\Phi+\ln \lambda \Psi).  \label{psh}
\ee
The fields $\Phi$ and $\Psi$ are what is called in two-dimensional conformal field theories a logarithmic pair \cite{Gurarie}. 
In the AdS literature is known as dipole pair \cite{fronsdal} and it is believed to describe singleton fields of the AdS group. 
This has also been confirmed in the AdS/CFT context \cite{ahmed,kogan}. It should be noted that 
Eqs. 
(\ref{psi}) and (\ref{phi}) reveal some problems. In fact, it is obvious that $\Phi$ satisfied the higher-order equation
\be
(\Box-m^2)^2\Phi=0
\ee
by acting with the Klein-Gordon operator on (\ref{phi}). This can also be seen, at the level of the action, by integrating out the 
$\Psi$ field in (\ref{slog}). Besides this apparent problem, logarithmic field theories seem to describe rather  successfully,
among others,  
percolation \cite{sale}, the quantum Hall effect \cite{QHE} as well as planar magnetohydrodynamics \cite{flo}.
As in the case of AdS, we expect that the ghost mode in this higher-derivative theory to 
be eliminated by appropriate gauge symmetry \cite{fronsdal,starinets}. Details will be given elsewhere.

The transformation (\ref{psh}) shows that, in general, 
two operators $\sigma_1$ and $\sigma_2$ of conformal dimension $w$ which under 
dilations transform as 
\be
i[D,\sigma_a]=\left(x^i\partial_i\delta_a^b+\Delta_a^b\right)\sigma_a, ~~~~a=1,2.
\ee
Up to now we have consider the case where the matrix $\Delta_a^b$ is diagonal and in 
particular $\Delta_a^b=w\,\delta_a^b$. However, we may consider a more general case
where ${\bf \Delta}=(\Delta_a^b)$ is brought to its  Jordan canonical form  
\be
{\bf \Delta}=\left(\begin{array}{cc}
                   w&0\\
                   1&w
                  \end{array} \right).
\ee
This means that $\sigma_a$ transforms under dilations $\vx\to \lambda \vx$ as 
\be
\sigma_a(\vx)\to \sigma'_a(\vx')=\Big{(}\exp[{\bf \Delta}\ln \lambda]\Big{)}{}_a^b\sigma_b(\lambda\vx)
\ee
and reproduce exactly the transformation (\ref{psh}) for $\Psi=\sigma_1\, ~\Phi=\sigma_2$.  
To find the correlators 
$G_{ab}(\vx,\vy)=\langle \sigma_\sigma^I(\vx)\sigma_\sigma^J(\vy)\rangle$ we 
may use the Ward identities for scale and special conformal transformations. Denoting by ${\bf G}$ the matrix $G_{ab}$,
scale invariance requires that ${\bf G}$ satisfies
\be
{\bf \Delta G}+{\bf G }{\bf \Delta}^T+r\frac{\partial}{\partial r}{\bf G}=0,\label{gs00}
\ee
where $r=|\vx-\vy|.$ Eq. (\ref{gs00})  is explicitly written as 
\be
&&\left(2 w+ x_{12}\frac{\partial}{\partial x_{12}}\right)G_{11}=0,  \nonumber \\
&&\left(2 w+ x_{12}\frac{\partial}{\partial x_{12}}\right)G_{12}+G_{11}=0,\nonumber \\
&&\left(2 w+ x_{12}\frac{\partial}{\partial x_{12}}\right)G_{22}+2G_{12}=0.
\label{gse}
\ee
In addition, special conformal transformation gives the constraint
\be
{\bf \Delta G}={\bf G}{\bf \Delta}^T,
\ee
which leads to
\be
G_{11}=0, ~~~~G_{12}=G_{21} .
\ee
We may then proceed to solve Eqs. (\ref{gse}), 
the solution of  which is provided by
\be
&&G_{12}= \frac{c}{|\vx-\vy|^{2w}}, ~~~~~~~~G_{11}=0, \nonumber \\
&&G_{22}= a \, G_{12}+\frac{\partial}{\partial w}G_{12}=\frac{c}{|\vx-\vy|^{2w}}
\left(-2 \ln |\vx-\vy|+a\right).
\ee
Therefore the two-point functions of the logarithmic pair $\sigma_1,\sigma_2$ turn out to be
\be
&&\langle \sigma_1(\vx)\sigma_2(\vy)\rangle=\langle \sigma_2(\vx)\sigma_1(\vy)\rangle= \frac{c}{|\vx-\vy|^{2w}}, 
\\
&&\langle \sigma_2(\vx)\sigma_2(\vy)\rangle=\frac{c}{|\vx-\vy|^{2w}}
\left(\phantom{\frac{1}{2}}\!\!\!\!\!\!-2 \ln |\vx-\vy|+a\right),  \\
&&\langle \sigma_1(\vx)\sigma_1(\vy)\rangle=0.
\ee
Let us now consider the three-point functions. Here we want to calculate the correlator 
\be
G_{abc}(\vx_1,\vx_2,\vx_3)=
\langle \sigma_a(\vx_1)\sigma_b(\vx_2)\sigma_c(\vx_3)\rangle\, .
\ee 
Again we will use 
Ward identities for dilations and special conformal 
transformations.
From dilation  we get 
\be
\Delta_a^iG_{ibc}+\Delta_b^iG_{aic}+\Delta_c^i G_{abi}+\left(\vx_1\cdot \vec{\nabla}_1 +\vx_2\cdot\vec{\nabla}_2 
+\vx_3\cdot\vec{\nabla}_3 \right)G_{abc}=0,
\ee
whereas from special conformal transformations we have ($\vec{b}$ being the parameter vector of the special conformal transformation, see Eq. (\ref{specconf}))

\be
&&\delta \vec{b}\cdot \left\{2 \vx_1\Delta_a^iG_{ibc}+2\vx_2\Delta_b^iG_{aic}+2\vx_3\Delta_c^i G_{abi}+\phantom{\frac{1}{X^X}}\right.\nonumber \\
&&\phantom{\frac{1}{X^X}} \left.\left[(\vx_1+\vx_2)
x_{12}\frac{\partial}{\partial x_{12}}+(\vx_1+\vx_3)
x_{13}\frac{\partial}{\partial x_{13}}+(\vx_2+\vx_3)
x_{23}\frac{\partial}{\partial x_{23}}\right]\right\}G_{abc}=0.
\ee
Combining the two equations above we get 
\be
\Delta_a^iG_{ibc}=\Delta_b^iG_{aic}=\Delta_c^i G_{abi},
\ee
which leads us to 
\be
&&wG_{122}+x_{ij}\frac{\partial}{\partial x_{ij }}G_{122}=0, ~~~~\forall\,  \,i<j  \label{3l1}\\
&&wG_{222}+G_{122}+x_{ij}\frac{\partial}{\partial x_{ij }}G_{122}=0, ~~~~\forall \, \, i<j     \label{3l2}\\
&&G_{111}=G_{112}=0.
\ee
The solution to Eqs.(\ref{3l1}) and (\ref{3l2}) is given by
\be
&&G_{122}=c x_{12}^{-w}x_{23}^{-w}x_{13}^{-w}, \\
&& G_{222}= c x_{12}^{-w}x_{23}^{-w}x_{13}^{-w}\left(2 \ln (x_{12}x_{23}x_{13})+a\right).
\ee
As a result, the three-point functions in the theory are given by
\be
&&\langle \sigma_1(\vx_1)\sigma_2(\vx_2)\sigma_2(\vx_3)\rangle=
\frac{c}{x_{12}^{w}x_{23}^{w}x_{13}^{w}},  \\
&&\langle \sigma_2(\vx_1)\sigma_2(\vx_2)\sigma_2(\vx_3)\rangle=\frac{c}{x_{12}^{w}x_{23}^{w}x_{13}^{w}}
\left\{\phantom{\frac{1}{2}}\!\!\!\!\!\!- \ln \big{(}x_{12}x_{23}x_{13}\big{)}+a\right\},  \\
&&\langle \sigma_1(\vx_1)\sigma_1(\vx_2)\sigma_2(\vx_3)=0,\\
&&\langle \sigma_1(\vx_1)\sigma_1(\vx_2)\sigma_1(\vx_3)=0.
\ee
The correlators in momentum space are easily evaluated. For example we find
\be \label{scc}
&&\langle \sigma_1(\vk_1)\sigma_2(\vk_2)\rangle'= \frac{C_0(w)}{k_1^{3-2w}}, \nonumber 
\\
&&\langle \sigma_2(\vk_1)\sigma_2(\vk_2)\rangle'=a\langle \sigma_1(\vk_1)\sigma_2(\vk_2)\rangle'+\frac{\partial}{\partial w}
\langle \sigma_1(\vk_1)\sigma_2(\vk_2)\rangle'= \frac{C_0(w)}{k_1^{3-2w}}\left(\phantom{\frac{1}{2}}\!\!\!\!\!\!-2 \ln k_1
+a+C_{0,w}\right), \nonumber \\
&&\langle \sigma_1(\vk_1)\sigma_1(\vk_2)\rangle=0,
\ee
where $C_{0,w}$ denotes  derivative of $C_0$ with respect to $w$. Similar expression holds for the three-point functions. 
For example we have that 
\be
\langle \sigma_2(\vk_1)\sigma_2(\vk_2)\sigma_2(\vk_3)\rangle'=a\langle \sigma_1(\vk_1)\sigma_2(\vk_2)\sigma_2(\vk_3)\rangle'+
\frac{\partial}{\partial w}
\langle \sigma_1(\vk_1)\sigma_2(\vk_2)\sigma_2(\vk_3)\rangle'.
\ee
In particular, 
in the squeezed $k_1\ll k_2,k_3$ we get 
\be
&&\langle \sigma_1(\vk_1)\sigma_2(\vk_2)\sigma_2(\vk_3)\rangle'\sim \frac{C_1(w)}{k_1^{3-w}k_2^{3-2w}}, ~~~\\
&&
\langle \sigma_2(\vk_1)\sigma_2(\vk_2)\sigma_2(\vk_3)\rangle'\sim \frac{C_1(w)}{k_1^{3-w}k_2^{3-2w}}
\left\{\phantom{\frac{1}{2}}\!\!\!\!\!\!\ln(k_1k_2^2)+a+C_{1,w}\right\},
\\
&&\langle \sigma_1(\vk_1)\sigma_1(\vk_2)\sigma_1(\vk_3)\rangle'=\langle \sigma_1(\vk_1)\sigma_2(\vk_2)\sigma_2(\vk_3)\rangle'=0.
\ee
The corresponding correlators of the comoving curvature perturbations can be easily calculated. For example, by 
using the expression (\ref{zeta}) we get for the 
spectrum
\be
\langle \zeta_{\vk_1}\zeta_{\vk_2}\rangle'\sim \frac{A}{k_1^{3-2w}}(1+2\gamma\ln k_1)\sim \frac{A}{k_1^{3-2w-2\gamma}}, \label{zs}
\ee
where 
\be
A=2N_1 N_2+aN_2^2 C_0+C_{0,w}, ~~~~\gamma=\frac{C_0}{N_2^2}{A}, 
\ee
and in the last step in (\ref{zs}) we have assumed that $\gamma\ll 1$. We see that the spectral index of the curvature perturbation power spectrum,  $n_\zeta-1={\rm d}\ln k^3 P^\zeta/{\rm d}\ln k$, gets a new contribution equal to $2\gamma$ from the 
 due to logarithmic short-distance singularities in the OPE, even if the fields involved are massless. 
 Further considerations will be presented elsewhere.

It should be noted that logarithmic field theories have OPE which contains short distance logarithmic singularities. For example
the OPE in Eq. (\ref{abvx}) is modified as \cite{Gurarie,Kogan}
\be
\sigma^I(\vx_1)\sigma^J(\vec{x}_2)\stackrel{\vx_1\to\vec{x}_2}{\sim}
\left(\frac{1}{x_{12}}\right)^{w_I\!+\!w_J}\left\{C_0^{IJ}+ 
D_0^{IJ}\ln|x_{12}|+ \frac{{C^{IJ}}_K}{x_{12}^{w_K}}\sigma^K(\vx_2)+ \cdots \right\}.
\ee
Repeating the analysis of the previous section, we can calculate the four-point function at the collapsed limit by using the above OPE 
in the (12) and (34) channels. The only difference is that thye matrix $C^{IJ}$ in Eq. (\ref{ccc}) is not positive definite. In fact,
a simple inspection of (\ref{scc}) reveals that, in the simplest case of two fields $(I,J=1,2)$, $C^{IJ}$ has a positive and  
a negative eigenvalues. Therefore, the Cauchy-Schwarz inequality (\ref{cn}) gets inverted and leads to 
\be
\tau_{\rm NL}\leq \left(\frac{6}{5}f_{\rm NL}\right)^2.
\ee 
Thus, logarithmic conformal field theories provide an example, consistent with the de Sitter symmetries,  which leads to violation
of the SY inequality. Such theories violate unitarity but there is no obvious reason for a CFT to be unitary \cite{strominger}, 
and logaritmic CFT's is an example. It remains to be seen if logarithmic conformal field theories play a real role in cosmology or not.

\section{Some considerations and conclusions}
\noindent
In this paper we have studied the implications of the symmetries present during a de Sitter phase for the statistical correlations of the light fields present during a multifield inflationary dynamics. In particular, we have assumed that the NG is generated by light fields other than the inflaton field. The cosmological perturbations are both scale invariant and conformally invariant. 

We have first shown  that, as a consequence of the conformal symmetries, the two-point cross-correlation of the light fields vanish if their conformal weights  are different. Therefore, no assumption is needed on such a cross-correlation, it is simply dictated by the conformal symmetry.

Secondly, we have pointed out that the OPE technique is very suitable to analyze two
 interesting limits: the squeezed limit of the three-point correlator and the collapsed limit of the four-point correlator. Despite the fact that the conformal symmetry does not fix the
 shape of the four-point correlators of the light NG fields, we have been able to compute it  
 in the collapsed limit. Both the resulting shapes of the squeezed limit of the bispectrum and the 
collapsed limit of the trispectrum  of the NG light fields turn out to be  of the same form of  the shapes of the corresponding bispectrum and trispectrum  universally generated on superhorizon scales of the  comoving curvature perturbation. Thanks to this result, we have succeeded 
in showing that the SY inequality relating the two NG observables $f_{\rm NL}$ and 
$\tau_{\rm NL}$ is valid independently of the NG nature of the light scalar fields at horizon crossing. In fact, we have been able to show that the SY inequality 
 is valid irrespectively of the conformal symmetry, being just a consequence of the 
short-distance expansion of the two-operator product expansion.

In most of this paper the working assumption was  that the cosmological perturbations enjoy both the scale invariant and the conformal symmetry of pure de Sitter. The  inflaton background
spontaneously breaks this symmetry, so that the variation of a correlation function of the curvature pertubation under the de
Sitter isometry group should always be connected with the soft emission of one or many soft  inflaton perturbations \cite{sloth1,sloth2,Creminelli2,by,sz}. It would be interesting to understand how our results will change under the assumption of a slight breaking of the de Sitter isometries. 
Under the assumption that the NG is generated by scalar fields other than the inflaton, we expect that our results in the squeezed and collapsed limits of the bispectrum and trispectrum respectively are still valid up to small corrections of the order of the slow-roll parameters.

Finally, while it is clear that
a  detection of a non-conformal correlation function, for example an
equilateral three-point function, would imply that the source of perturbations is not decoupled from the inflaton \cite{Creminelli1}, it would be interesting to understand if it possible to find other cosmological observables which can robustly test the conformality of the  primordial cosmological perturbations. This might be a non trivial task as  post inflationary nonlinear evolution of the correlators  contaminate such a primordial input.

\section*{Acknowledgments}

We thank  C. Byrnes, P. Creminelli, A. Petkou and M. Sloth for useful conversations. A.R. is supported by the Swiss National
Science Foundation (SNSF), project `The non-Gaussian Universe" (project number: 200021140236).


\end{document}